\PassOptionsToPackage{table,xcdraw}{xcolor}
\documentclass[10pt,conference]{IEEEtran}
\usepackage[table]{xcolor}  
\ifCLASSOPTIONcompsoc
  \usepackage[nocompress]{cite}
\else
  \usepackage{cite}
\fi
\ifCLASSINFOpdf
\else
\fi
\usepackage{amsmath}
\usepackage{balance}  
\usepackage{graphics} 
\usepackage{txfonts}
\usepackage{times}    
\usepackage{color}
\usepackage{textcomp}
\usepackage{booktabs}
\usepackage{ccicons}

\usepackage{cite}
\usepackage{url}
\usepackage{fancybox}
\usepackage{multirow}
\usepackage{flushend}
\usepackage{booktabs}
\usepackage{tabularx}
\usepackage{comment}
\usepackage{array}
\usepackage[flushleft]{threeparttable}
\usepackage{mdframed}
\graphicspath{{}{images/}{dia/}}
\DeclareGraphicsExtensions{.pdf,.png}

\usepackage{listings}
\usepackage{courier}
\usepackage{hyperref}

\usepackage{xpatch}

\xpatchcmd{\refstepcounter}{%
  \stepcounter{#1}%
}{%
  \stepcounter{#1}%
  \xdef\scr{\number\value{#1}}%
}{\typeout{success}}{\typeout{failure}}

\newcounter{o}
\usepackage{tikz}
\usepackage{styles/pgf-pie}
\usetikzlibrary{positioning,shadows}
\usepackage{balance}

\newif\ifpienumberinlegend
\pgfkeys{/number in legend/.code=
    \expandafter\let\expandafter\ifpienumberinlegend
    \csname if#1\endcsname
    \ifpienumberinlegend

    \def\beforenumber##1\afternumber{}%
    \fi,
    /number in legend/.default=true
}
\definecolor{1c1}{RGB}{188,162,6}
\definecolor{1c2}{RGB}{137,129,80}
\definecolor{1c3}{RGB}{239,167,31}
\definecolor{1c4}{RGB}{88,194,241}
\definecolor{1c5}{RGB}{6,180,188}

\tikzset{mynode/.style={draw=white,solid,circle,fill=green,inner sep=1pt, thick,
text=black}}
\tikzset{arrow line/.style={dashed, line width= 2.5pt, color=#1}}

\def\bf{\textbf}

\def\fig {Fig.~}

\def\tbl {Table~}

\def\sec {Section~}
\def\secs {Sections~}

\def\it{\textit}

\def\tt{\mct}

\newcommand{\mct}[1]{{\footnotesize {\texttt {#1}}}}

\usepackage{paralist}

\usepackage{tikz}
\newcommand*\circled[1]{\tikz[baseline=(char.base)]{
            \node[shape=circle,draw,inner sep=1pt] (char) {#1};}}

\lstdefinestyle{inlinecode}{basicstyle={\ttfamily\scriptsize\bfseries}}

\newcommand{\urls}[1]{{\scriptsize\url{#1}}}
\usepackage{tcolorbox}
\newcommand{\emt}[1]{\emph{``#1''}}

\usepackage{paralist}
\usepackage[outercaption]{sidecap}
\usepackage [autostyle, english = american]{csquotes}
\MakeOuterQuote{"}
\newcounter{scn}
\usepackage[shortlabels]{enumitem}
\usepackage{bchart}

\newcounter{finding_counter}
\newcommand{\threebars}[3]{
{{\color{black}\rule{#1pt}{4pt}} C #1\%}
{{\color{magenta}\rule{#2pt}{4pt}} R #2\%}
{{\color{green}\rule{#3pt}{4pt}} O #3\%}
}

\newtoggle{comment}
\toggletrue{comment}

\usepackage{tcolorbox}
\newlist{RQ}{enumerate}{1}
\setlist[RQ, 1]{label = RQ \arabic*:}

\usepackage{forest}
\usepackage{soul}

\usepackage[multiple]{footmisc}
\usepackage{threeparttable} 

\usepackage{subcaption}
\usepackage{graphicx}

\newcommand{\TDstat}[3]{
{{\color{black}\rule{#1pt}{4pt}}} {{#2} Comments}
{{\color{magenta}\rule{#3pt}{4pt}} {#3}\% Comments}
}

\begin{document}
\title{Automatic Detection and Analysis of Technical Debts in Peer-Review Documentation of R Packages}

\author{
\IEEEauthorblockN{Junaed Younus Khan and Gias Uddin\\DISA Lab, University of Calgary}
}













\IEEEtitleabstractindextext{%
\begin{abstract}
Technical debt (TD) is a metaphor for code-related problems that arise as a result of prioritizing speedy delivery over perfect code. Given that the reduction of TDs can have long-term positive impact in the software engineering life-cycle (SDLC), TDs are studied extensively in the literature. However, very few of the existing research focused on the technical debts of R programming language despite its popularity and usage. Recent research by Codabux et al. \cite{codabux2021technical} finds that R packages can have 10 diverse TD types analyzing peer-review documentation. However, the findings are based on the manual analysis of a small sample of R package review comments. In this paper, we develop a suite of Machine Learning (ML) classifiers to detect the 10 TDs automatically. The best performing classifier is based on the deep ML model BERT, which achieves F1-scores of 0.71 - 0.91. We then apply the trained BERT models on all available peer-review issue comments from two platforms, rOpenSci and BioConductor (13.5K review comments coming from a total of 1297 R packages). We conduct an empirical study on the prevalence and evolution of 10 TDs in the two R platforms.  We discovered documentation debt is the most prevalent among all types of TD, and it is also expanding rapidly. We also find that R packages of generic platform (i.e. rOpenSci) are more prone to TD compared to domain-specific platform (i.e. BioConductor).  Our empirical study findings can guide future improvements opportunities in R package documentation. Our ML models can be used to automatically monitor the prevalence and evolution of TDs in R package documentation.

\end{abstract}

\begin{IEEEkeywords}
R, Technical Debt, Machine Learning, Empirical Study, rOpenSci, BioConductor, Documentation
\end{IEEEkeywords}}

%


\maketitle

\IEEEdisplaynontitleabstractindextext
\section{Introduction}\label{sec:introduction}


Technical debt (TD) denotes sub-optimal software development choices/actions \cite{cunningham1992wycash, avgeriou2016managing}. In order to meet project goals with time and resource constraints, developers often use shortcuts or execute rapid hacks in their work \cite{besker2018embracing, seaman2011measuring}. While such shortcuts may enable developers in achieving their short-term objectives, they may have harmful long-term consequences \cite{kruchten2013technical, lim2012balancing}. Studies find that software applications containing TDs are more prone to issues like bugs and maintenance problems \cite{fernandez2014guiding, kruchten2012technical, li2015architectural, sultana2020examining, hall2014some}. TDs also hinder the productivity of software developers \cite{besker2018technical}. Several studies have been conducted to understand the pattern and impact of different types of TD. Most of the existing studies used a single artifact (i.e. source code comments) to explore TD with a few recent exceptions \cite{alves2016identification, bellomo2016got, xiao2016identifying}. 

R is a popular programming language among data scientists and scientific communities. While studies in TDs have mainly focused on popular programming languages like Java, we are aware of few research that focused on TDs in R packages. Recently, Codabux et al. \cite{codabux2021technical} manually analyzed the peer-review documentation explore TD in R \href{https://ropensci.org/}{rOpenSci} packages. rOpenSci is a popular platform that promotes development and use of high-quality R software \cite{boettiger2015building}. It conducts peer-review of R packages in public GitHub issues. Codabux et al. \cite{codabux2021technical} identified 10 types of TD in the a sample of 600 comments related to rOpenSci R packages. They are documentation, code, design, defect, requirement, test, architecture, build, usability, and versioning debts. However, to get a complete understanding of the distribution and growth of different types of TD in different R packages, a large-scale empirical study is necessary. Codabux et al. also focused on only one platform (rOpenSci). To get a more representative picture of TDs in R packages, other platforms of R packages should be studied too. Without an automatic detection of TDs in R packages, it is however, not possible to facilitate such large-scale analysis of TDs in packages. This is because manual labeling is a time-consuming and resource-intensive process. Automatic detection is also crucial to take necessary precautions about TD, and to manage and fix them. Most existing researches basically focused on automatic detection of self-admitted technical debts (SATD) using source-code \cite{da2017using, maldonado2017empirical, yan2018automating, potdar2014exploratory, sierra2019survey}. As such, we are aware of no research that automatically detected the 10 TD types from Codabux et al. \cite{codabux2021technical} in R packages.

In this paper, we develop techniques to automatically detect 10 TD types from Codabux et al. \cite{codabux2021technical} in R packages. We then conduct an empirical study to understand the prevalence and evolution of the 10 TDs in all R packages from rOpenSci \cite{boettiger2015building} and BioConductor \cite{gentleman2004bioconductor}. Specifically, we follow three three major phases (see \fig\ref{fig:schematic_digram}). First, we employed a two-stage BERT-based hierarchical machine learning framework to detect TD types. Second, we apply the trained model on all approved comments from the peer review package documentation of two R platforms, rOpenSci and BioConductor. Third, we analyze the prevalence of evolution of the 10 TD types across the two R platforms based on the data labeled by the trained BERT model. To the end, we answer three research questions:
\begin{figure}[t]
  \centering
  \includegraphics[scale=.6]{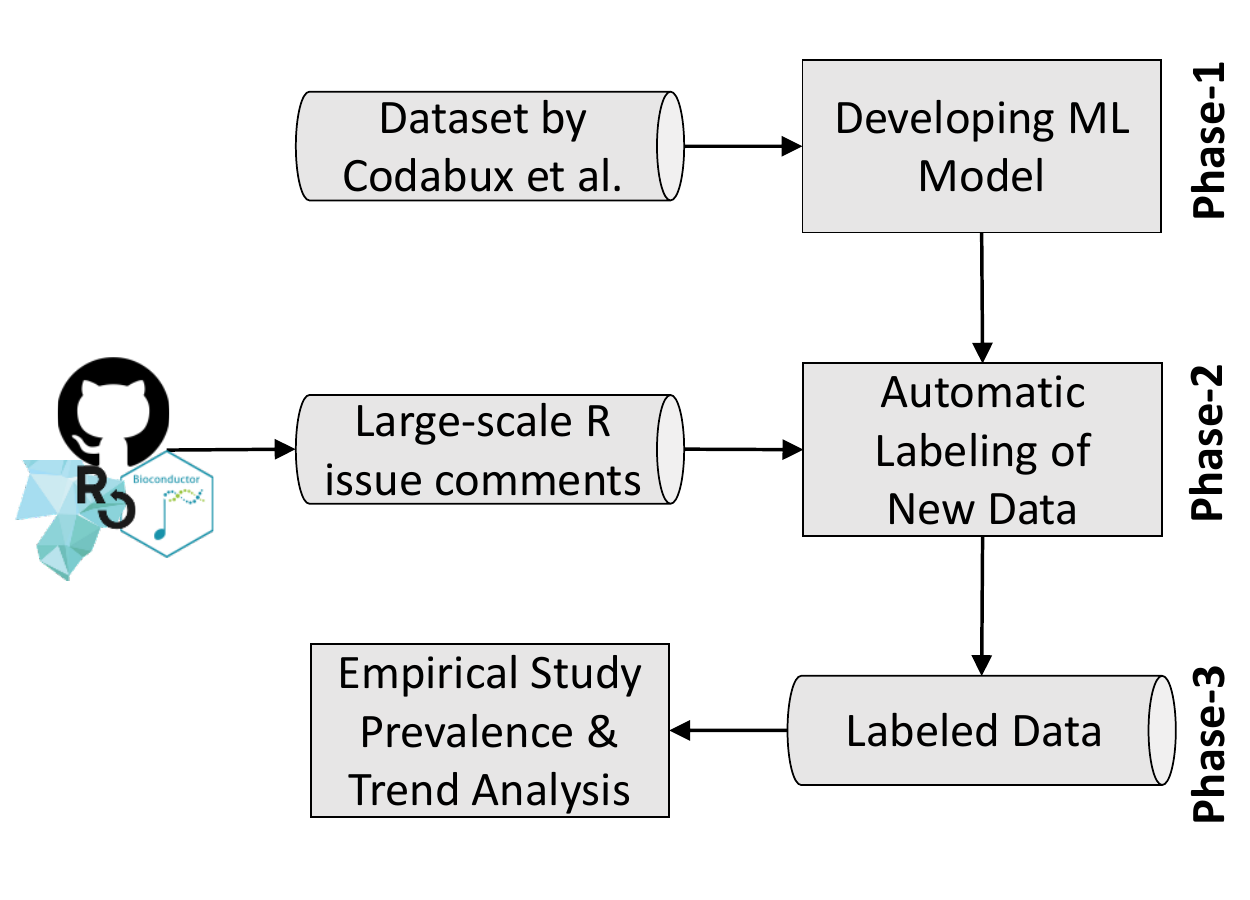}
  \caption{Schematic diagram of our study.}

  \label{fig:schematic_digram}
\vspace{-1mm}
\end{figure}

\textbf{RQ1. How accurately can we detect TDs in R packages? (\sec\ref{sec:rq1})} 
First, we checked whether a given sentence (of issue comments) indicates any TD or not, if yes, we determined the type of that TD in the second stage. We evaluated a suite of ML models (i.e. SVM, BiLSTM, and BERT) to detect the presence of TD in the first stage. The BERT-based model outperforms the other two with an F1-score of 0.90 in this task. Second, we detected the types of these TD using a BERT-based hierarchical approach which shows F1-scores between 0.71 -- 0.91 to detect the 10 TDs.

\textbf{RQ2. How prevalent are TDs in R packages? (\sec\ref{sec:rq2})} We applied our BERT based framework on the peer-review issue comments of all approved rOpenSci and BioConductor packages. It is important to know the current state of different type of TD e.g. which type is more (or less) frequent than others. Hence, we analyzed the distribution of different types of TDs across all rOpenSci and BioConductor packages. We find that documentation debt is the most prevalent, while versioning debt is the least frequent.      

\textbf{RQ3. How do the TDs evolve in R packages? (\sec\ref{sec:rq3})} Trend analysis shows us the growing pattern of different TD and helps us to understand what to expect in the future. We analyzed the evolution of different TDs over time and observed that there is a significant upward trend of most of the TD (specially documentation, defect, test debts) in rOpenSci. On the other hand, there is an overall decline of all TD types in BioConductor packages.

To assist R developers with our research, we developed a browser extension that can show the TD types of a given R package in rOpenSci and BioConductor. The browser extension uses our developed ML model to automatically detect the TD types of an R package. Our developed ML model and the browser extension can support R developers and researchers to learn and to conduct empirical studies of the 10 TD types in R packages. We are aware of no previous empirical study that analyzed all R packages in rOpenSci and BioConductor. As such, our tool and empirical study findings can benefit diverse stakeholders like R developers, package maintainers, and researchers (see \sec\ref{sec:implications}).

\textbf{Replication Package.} Our code and data are shared at \url{https://github.com/disa-lab/R-TD-SANER2022}

\section{Automatic Detection of R TDs (RQ1)}\label{sec:rq1}
In this section, we describe the design and performance of a suite of Machine Learning (ML) models to automatically detect 10 TD types in R packages. Automatic detection is important to monitor, analyze and to fix TDs.    

\subsection{Benchmark Dataset Used to Design ML Models} 
To design and test the TD type detection algorithms, we used the dataset published by Codabux et al. \cite{codabux2021technical} which contains 600 TD instances in R peer-reviews documentation of rOpenSci manually labelled in 10 TD types. Each instance is a phrase extracted from a full issue comment. First, they extracted the comments of all 157 approved rOpenSci packages of that time. Second, they manually investigated a randomly selected subset of the comments and identified the ones with potential TD indication. Later, 600 phrases were manually extracted from those comments and labelled in different TD types. Hence, one limitation of the dataset is that it only contains TD significant parts of the issue comments and remaining parts of the comments (i.e. insignificant/non-TD sentences) were discarded. Since issue comments are from natural human language domain, it is obvious that they contain a number of sentences that are not relevant to TD. Hence, for automatic detection, we need a number of such non-TD sentences as well for training our machine learning model so that it can distinguish them from TD sentences and discard them from a given issue comment in real world scenario. Accordingly, we built a dataset of both TD-significant and TD-insignificant sentences as follows. \begin{inparaenum}[(1)]
\item We extracted the full comments of TD instances labeled in the dataset \cite{codabux2021technical} using their comment-ids. \item We split the extracted comments into sentences. \item Among these sentences, ones that exist in the TD dataset of \cite{codabux2021technical} are considered significant for determining TD. \item The rest sentences are considered insignificant. In this process, we obtained a binary dataset of total of 1205 sentences: 805 TD significant and 400 TD insignificant (i.e. non-TD). Each of these 805 TD sentences again carries a type (i.e. TD type) as labeled in the original dataset \cite{codabux2021technical}. \end{inparaenum}

\begin{table}
\centering
\caption{Statistics of the Benchmark Dataset}
\scalebox{.9}
{
\begin{tabular}{@{}lccl@{}}
\toprule
\textbf{TD Type} & \textbf{\# Sentence} & \bf{\#Word/Sent} & \textbf{Top 3 Words}         \\ \midrule
N/A              & 400                  & 9                                                                              & thanks, package, CRAN           \\
Documentation    & 230                  & 11                                                                             & documentation, vignette, readme \\
Code             & 226                  & 9                                                                              & function, name, variable        \\
Design           & 202                  & 11                                                                             & function, data, method          \\
Defect           & 192                  & 11                                                                             & error, code, data               \\
Requirement      & 207                  & 12                                                                             & user, data, think               \\
Test             & 195                  & 11                                                                             & test, coverage, full            \\
Architecture     & 200                  & 12                                                                             & package, make, tool             \\
Build            & 172                  & 10                                                                             & check, install, CRAN            \\
Usability        & 150                  & 13                                                                             & error, help, message            \\
Versioning       & 104                  & 12                                                                             & version, current, numbers       \\ \bottomrule
\end{tabular}
}
\label{dataset_table_1}
\end{table}



One issue of the dataset is it is greatly imbalanced. 
Although as claimed by the authors \cite{codabux2021technical}, it might represent the comparative distribution of different types of debts in their studied subset of R domain, we need a slightly more balanced dataset to develop an unbiased machine learning model. We augmented the text data by replacing words (adjectives, verbs) with their synonyms to generate different texts with the same meaning \cite{synonym_replace_text_aug_2,synonym_replace_text_aug_3}. The synonyms were generated using WordNet \cite{miller1995wordnet}. We did not use SMOTE, since it is not effective for the high dimensional numerical representation of text data \cite{smote_prob_for_text}. Table \ref{dataset_table_1} shows statistics of the final dataset.

\subsection{The TD Detection Algorithms}
\begin{figure}[t]
  \centering
  \includegraphics[scale=0.5]{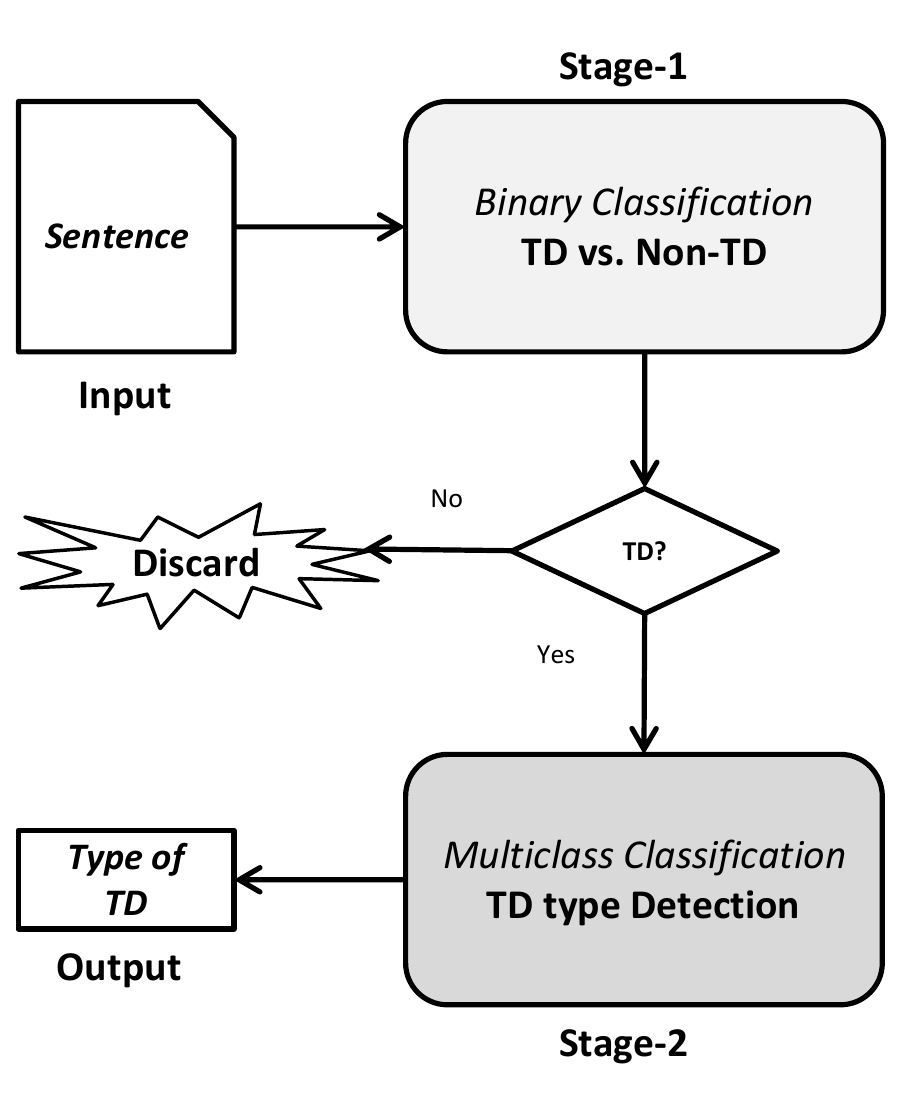}
  \caption{Automatic Detection of different types of TD.}

  \label{fig:td_type_detection_workflow}
\vspace{-1mm}
\end{figure}
We developed a two-stage framework to detect TD instances of a package from corresponding GitHub issue comments as depicted in Figure \ref{fig:td_type_detection_workflow}. For a given issue comment, we split it into sentences and every sentence is labeled as TD significant or insignificant in stage-1. TD insignificant sentences are discarded and the TD significant sentences are passed to stage-2 for TD-type detection. The detailed workflow of the two stages are mentioned below.

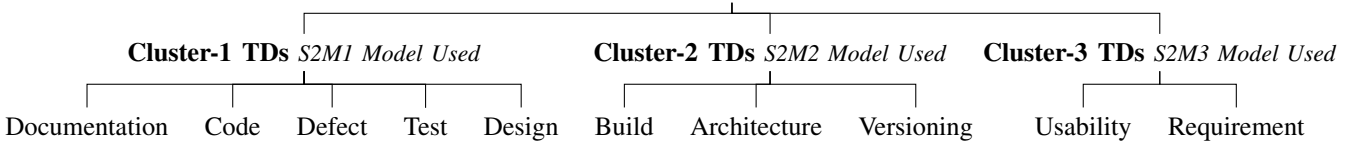
\begin{figure*}
\centering
\begin{forest}
for tree={%
    edge path={\noexpand\path[\forestoption{edge}] (\forestOve{\forestove{@parent}}{name}.parent anchor) -- +(0,-12pt)-| (\forestove{name}.child anchor)\forestoption{edge label};}
}
[
\bf{TD Type Clusters} \small{\it{S2M0 Model Used for Cluster Detection}}
[
\bf{Cluster-1 TDs} \it{\small{S2M1 Model Used}}
[Documentation]
[Code]
[Defect]
[Test]
[Design]
]
[
\bf{Cluster-2 TDs} \small{\it{S2M2 Model Used}}
[Build]
[Architecture]
[Versioning]
]
[
\bf{Cluster-3 TDs} \small{\it{S2M3 Model Used}}
[Usability]
[Requirement]
]
]
\end{forest}

\caption{Hierarchy of different types of TD}
\label{TD_Hierarchy_fig}
\end{figure*}

\subsubsection{Stage 1. Discarding Insignificant Sentences}
We developed a BERT-based model that would detect TD-significant vs TD-insignificant instances in a binary classification fashion. BERT is a pre-trained model which was designed to learn contextual word representations of unlabeled texts \cite{Delvin-BERTArch-Arxiv2018}. We chose BERT because it is found to significantly outperform other models in various natural language processing and text classification tasks \cite{bert_success_1,bert_success_2,bert_success_5}. We used BERT-Base containing 12 layers with 12 attention heads and 110 M parameters. We appended a classification head composed of a single linear layer on top of it. We trained the model for 10 epochs with a mini-batch size of 32 and maximum input length of 300. We used early-stop to avoid overfitting
\cite{early_stop_bert_1} considering validation loss as the metric of the
early-stopping \cite{early_stop_bert_2}. We used AdamW optimizer \cite{adam_w} setting the learning
rate to 4e\textsuperscript{-5}, $\beta1$ to 0.9, $\beta2$ to 0.999, and $\epsilon$ to
1e\textsuperscript{-8}
\cite{Delvin-BERTArch-Arxiv2018, bert_fine_tuning}. We used binary cross-entropy to calculate the loss \cite{binary_cross_AreLossFunctionSame}. We refer to this model as \textbf{S1M (Stage-1 Model)} for the rest of this paper. We compared the performance of this BERT-based model (S1M) with one traditional ML model (i.e. SVM) and one deep learning model (i.e. Bi-LSTM). SVM is widely used for text classification \cite{dumais1998using, colas2006comparison}. We used linear kernel for the SVM as prescribed by earlier works \cite{zhang2008text} with Bag of words (BoW). BoW is a popular feature extraction approach for text data and successfully applied for text classification \cite{bag_of_words_for_text_1}. On the other hand, Bi-LSTM is capable of exploiting contextual information from text data examining from both directions \cite{graves2005framewise}. For Bi-LSTM model, we used 300 hidden states with ADAM optimizer \cite{kingma2014adam} and an initial learning rate of 0.001. We used 100 dimensional pre-trained GloVe  embedding as input features \cite{pennington2014glove} and trained the model with 256 batch-size over 10 epochs.    

\subsubsection{Stage 2. Detection of TD-types in Significant Sentences}
In this stage, we want to detect the type of TD from a given instance (i.e. sentence). We consider ten types of TD discussed in \cite{codabux2021technical}. Hence, it is a 10-class classification problem. Since BERT was the best performing model for TD presence detection in the first stage (see Section \ref{subsubsection:stage1_evaluation_results}), we limited our experiment to BERT for this stage. We first evaluated the effectiveness of a single classifier to decide among all the 10 possible classes. However, the single BERT performed poorly with an accuracy of $\sim$12\% which is nearly random for a 10-class problem. In the multi-class scenario, the classification task becomes very difficult since the classifier has to distinguish between a large number of classes in order to make predictions. To deal with this issue, we leveraged the concept of hierarchical classification \cite{arabie1996hierarchical,silla2011survey}. Hierarchical classification approaches the multi-class problem by splitting the output space like a tree where each (parent) node is divided into a number of child nodes, and the procedure is repeated until each child node represents a single class \cite{chaitra2018review}. Hence, it needs a predefined data taxonomy according to which it hierarchically distributes all the classes in a collection of multi-class sub-problems and thus the number of classes involved in each local sub-problem gets reduced. To induce the class hierarchy (i.e. taxonomy), we followed the procedure proposed by Daniel et al. \cite{confusionMatrix_for_hierarchy}. We used the $10\times10$ confusion matrix produced while using a single classifier to predict among all 10 classes together. First, we generated a distance matrix $(D)$ from the confusion matrix $(M)$ using Equations \ref{equation:distance_matrix_step1} and \ref{equation:distance_matrix_step2}. 
\begin{equation}
\label{equation:distance_matrix_step1}
\textrm{Confusion matrix normalization}~\overline{M}(i,j) = \frac{M_{ij}}{\sum_{j=1}^{n}M_{ij}}
\end{equation}
\begin{equation}
\label{equation:distance_matrix_step2}
\textrm{Distance Matrix}~D(i,j) = \left\{ 
  \begin{array}{ c l }
    1- \frac{\overline{M}_{ij}+\overline{M}_{ji}}{2} & \quad \textrm{if } i \neq j \\
    0                 & \quad \textrm{if } i = j
  \end{array}
\right.
\end{equation}

Any element of the distance matrix, $D(i,j)$ indicates the level of similarity between the class $i$ and $j$ ranging from 0 to 1 where 0 (and 1) means completely indistinguishable (and distinguishable). Then we divided all 10 types of TD in 3 clusters using spectral clustering algorithm \cite{spectralCluster1, spectralCluster2}. The optimal number of cluster, K (=3) was determined using eigengap heuristic \cite{spectralCluster2}. Thus, we induced a hierarchy of TD types as showed in Figure \ref{TD_Hierarchy_fig}. Then we used a total of 4 BERT-based classifiers (each for every parent node) where the first classifier (of the root node) was trained to distinguish among the three clusters only. To achieve this goal, we relabeled all the data of our dataset into three classes according to the hierarchy of Figure \ref{TD_Hierarchy_fig} and trained the first classifier on it. We call it \textbf{S2M0 (Stage-2 Model-0)}. Similarly, the next three classifiers (addressed later as \textbf{S2M1, S2M2, S2M3}) were trained to detect debts of Cluster-1, Cluster-2, and Cluster-3 respectively. Hence, \textbf{S2M1} detects among documentation, code, defect, test, and design debts; \textbf{S2M2} detects among build, architecture, and versioning debts; \textbf{S2M3} detects among usability and requirement debts.

\subsection{Evaluation of the ML Models}
For the first stage, we used our dataset in binary fashion (i.e. consisting two labels: TD and non-TD). For the second stage, we only considered the TD significant sentences labelled in 10 types (since the non-TD sentences will be eliminated in the first stage). We split the dataset in 80:20 ratio to create train and test sets in every stage of classification. 


We report the classification performance using four standard metrics: Accuracy, Precision, Recall, F1-score \cite{Manning-IRIntroBook-Cambridge2009}. Accuracy ($A$) is the ratio of the number of 
correctly predicted examples out of all data. Precision ($P$) is the ratio of the number of correctly predicted examples and all the predicted examples for a given class.
Recall ($R$) is  the ratio of the number of correctly predicted examples and  all  examples of a given class. F1-score ($F1$) is the harmonic mean of
precision and recall.

{\scriptsize
\begin{eqnarray*}
P  = \frac{TP}{TP+FP},~
R = \frac{TP}{TP+FN},~
F1 = 2*\frac{P*R}{P+R},~
A = \frac{TP+TN}{TP+FP+TN+FN}
\end{eqnarray*}}
TP = Correctly classified as a TD, 
FP = Incorrectly classified as a TD, TN = Correctly classified as not a TD, 
FN = Incorrectly classified as not a TD. 

\subsubsection{Detection Performance of Significant vs non-Significant Comments for TD}
\label{subsubsection:stage1_evaluation_results}
In stage-1, we developed a BERT based binary classifier (S1M) for discarding TD insignificant sentences from an issue comment. We also evaluated SVM and Bi-LSTM for these tasks as baselines. We report the performance of stage-1 in Table \ref{table_model_performance_stage1}. We can see that our BERT-based model (S1M) outperforms both SVM and Bi-LSTM with 91\% accuracy and 90\% F1-score. Since the task of stage-1 (i.e. binary classification) is relatively simple and straight-forward, state-of-the-art pre-trained language model like BERT shows very high performance here \cite{bert_success_1, bert_success_2}.


\subsubsection{Detection Performance of TD Types in Significant Comments}
 We report the result of stage-2 in Table \ref{table_model_performance_stage2}. We see that hierarchical classification greatly outperforms the single-model approach. In fact, that was our motivation behind employing hierarchical approach. Among the hierarchical classifiers (of different stages), S2M1 has to distinguish among 5 classes of cluster-1 (i.e.documentation, code, defect, test, design debts). It achieves 72\% accuracy with 71\% F1-score. Other models of stage-2 (i.e. S2M0, S2M2, S2M3) show better performance with F1-scores of 81\%, 77\%, and 91\% respectively, since they work with comparatively smaller clusters. Specially, cluster-3 contains only 2 classes (i.e. usability, requirement debts), hence, the classification problem becomes binary here and S2M3 achieves very high performance.        


\begin{table}
\centering
\caption{Performance of Stage-1: TD Presence Detection}
\begin{tabular}{@{}lrcccc@{}}
\toprule
\textbf{Feature} & \textbf{Model} & \textbf{A}    & \textbf{P}    & \textbf{R}    & \textbf{F}    \\ \midrule
\textbf{BoW}              & \textbf{SVM}            & 0.76          & 0.77          & 0.76          & 0.76          \\ 
\textbf{GloVe Embed}      & \textbf{Bi-LSTM}        & 0.82          & 0.82          & 0.82          & 0.82          \\ 
\textbf{BERT Embed}       & \textbf{BERT (S1M)}    & {0.91} & {0.90} & {0.90} & {0.90} \\ \bottomrule
\end{tabular}
\label{table_model_performance_stage1}
\end{table}

\begin{table}
\centering
\caption{Performance of Stage-2: TD Type Detection}
\resizebox{\columnwidth}{!}
{
\begin{tabular}{llcccc} 
\toprule
\textbf{Model}        & \textbf{Task}          & \textbf{A} & \textbf{P}               & \textbf{R}               & \textbf{F}                \\ 
\midrule
\textbf{Single BERT } & All 10 types TD detection    & 0.12       & 0.07                     & 0.08                     & 0.08                      \\ 
\midrule
\multicolumn{6}{l}{\textbf{Hierarchical BERT to Detect the 10 TDs in Different Clusters}}                                                                                                \\ 

\textbf{S2M0}         & Cluster detection      & 0.86       & 0.82                     & 0.82                     & 0.82                      \\
\textbf{S2M1}         & Cluster-1 TD detection & 0.72       & 0.73                     & 0.70                     & 0.71                      \\
\textbf{S2M2}         & Cluster-2 TD detection & 0.77       & \multicolumn{1}{l}{0.77} & \multicolumn{1}{l}{0.76} & \multicolumn{1}{l}{0.77}  \\
\textbf{S2M3}         & Cluster-3 TD detection & 0.93       & \multicolumn{1}{l}{0.93} & \multicolumn{1}{l}{0.89} & \multicolumn{1}{l}{0.91}  \\
\bottomrule
\end{tabular}
}
\label{table_model_performance_stage2}
\end{table}

\subsubsection{Misclassification Analysis}
We further analyzed what factors the single-BERT made perform poorly and how they were solved by our hierarchical-BERT approach. Table \ref{table_misclassification} shows some examples of TD instances that were misclassified by single-BERT but correctly classified by the hierarchical approach. We find that the most probable reason behind single-BERT's misclassification is the confusion among the classes within the same cluster (see Figure \ref{TD_Hierarchy_fig}). In fact, that was the intuition behind our hierarchical classification. Additionally, a few misclassification occurred due to lack of context and vocabulary mismatch as well. For example, ``Though there are tests to cover most functions, good-practice still flags some lines as not being covered.'' was misclassified as code debt by Single-BERT while it is actually indicating a test debt due to the presence of common words across TDS like `functions', `flag', `good-practice', `lines', etc.

\begin{table*}
\centering
\rowcolors{2}{}{lightgray!30}
\caption{Misclassification Analysis of Single-BERT where Hierarchical BERT Correctly Classified}
\resizebox{\textwidth}{!}
{
\begin{threeparttable}
\begin{tabular}{llll} 
\toprule
\textbf{TD Instance Example Comment}       & \textbf{Actual TD} & \textbf{Single-BERT Label} & \bf{Misclassification Reason}  \\ 
\midrule
\textit{``With the data I would prefer to keep the internal df in the sysdata.rda file.''\href{https://github.com/ropensci/software-review/issues/284}{\textit{ I\textsubscript{r284}}}\tnote{*}}
& Architecture~~                                                                                                            & Build                                                                                 & Lack of context                                                                                  \\ 
\begin{tabular}[c]{@{}l@{}}\textit{``I ran into a small problem when trying to install via CRAN which I reported} \\ \textit{an issue.'' \href{https://github.com/ropensci/software-review/issues/115}{\textit{ I\textsubscript{r115}}}}\end{tabular}                                                & Build                                                                                                                     & Documentation                                                                         & Lack of context                                                                                  \\ 
\begin{tabular}[c]{@{}l@{}}`\textit{`There are still some unconventional R styles used that hindered my ability}\\\textit{to understand the package.'' \href{https://github.com/ropensci/software-review/issues/139}{\textit{ I\textsubscript{r139}}}}\end{tabular}                                    & Code                                                                                                                      & Design                                                                                & Confusion within Cluster-1                                                                       \\ 
\textit{``There is a bug in how double last names (as in Spanish) are generated.'' \href{https://github.com/ropensci/software-review/issues/94}{\textit{ I\textsubscript{r94}}}}                                                                                                            & Defect                                                                                                                    & Code                                                                                  & Confusion within Cluster-1                                                                       \\ 
\begin{tabular}[c]{@{}l@{}}\textit{``But I think making some commonly used flags into first-class arguments} \\\textit{would dramatically improve the appeal of this package. \href{https://github.com/ropensci/software-review/issues/139}{\textit{ I\textsubscript{r139}}}''}\end{tabular}           & Design                                                                                                                    & Build                                                                                 & Lack of context                                                                                  \\ 
\textit{``However, the text of the vignette is clearly still unfinished.'' \href{https://github.com/ropensci/software-review/issues/94}{\textit{ I\textsubscript{r94}}}}                                                                                                                    & Documentation                                                                                                             & Defect                                                                                & Confusion within Cluster-1                                                                       \\ 
\begin{tabular}[c]{@{}l@{}}\textit{``I recommend removing the flexibility and only letting the API receive JSON,}\\\textit{since validation information is only returned for JSON data.'' \href{https://github.com/ropensci/software-review/issues/141}{\textit{ I\textsubscript{r141}}}}\end{tabular} & Requirement                                                                                                               & Usability                                                                             & Confusion within Cluster-3                                                                       \\ 
\begin{tabular}[c]{@{}l@{}}\textit{``Though there are tests to cover most functions, good-practice still flags} \\\textit{some lines as not being covered.'' \href{https://github.com/ropensci/software-review/issues/290}{\textit{ I\textsubscript{r290}}}}\end{tabular}                               & Test                                                                                                                      & Code                                                                                  & Vocabulary mismatch                                                                              \\ 
\begin{tabular}[c]{@{}l@{}}\textit{``After that I'll focus on some of the bigger changes for a v0.3.0 release, since}\\\textit{they'll likely introduce some breaking changes.'' \href{https://github.com/ropensci/software-review/issues/279}{\textit{ I\textsubscript{r279}}}}\end{tabular}          & Versioning                                                                                                                & Build                                                                                 & Confusion within Cluster-2                                                                       \\ 
\begin{tabular}[c]{@{}l@{}}\textit{``You'd have a much more welcoming package that doesn't have to direct its} \\\textit{users to a GNU manpage in the opening paras of its vignette!'' \href{https://github.com/ropensci/software-review/issues/139}{\textit{ I\textsubscript{r139}}}}\end{tabular}   & Usability                                                                                                                 & Requirement                                                                           & Confusion within Cluster-3                                                                       \\
\bottomrule
\end{tabular}
\begin{tablenotes}\footnotesize
\item[*] Link to corresponding rOpenSci issue \textit{(I\textsubscript{rx})}
\end{tablenotes}
\end{threeparttable}
}
\label{table_misclassification}
\end{table*}

\section{Empirical Study of R TDs}
In this section, we report an empirical study by applying the best performing model from RQ1 (i.e., Hierarchical BERT) on all issue comments from R packages of two platforms, rOpenSci and BioConductor. We answer two research questions:
\begin{enumerate}[start = 2, label=\bf{RQ}\arabic{*}., leftmargin=28pt]
    \item How prevalent are the 10 TD types in the R packages across the two platforms? (\sec\ref{sec:rq2})
    \item How do the 10 TD types evolve in the studied R packages? (\sec\ref{sec:rq3})
\end{enumerate} The prevalence analysis of TDs (RQ2) can inform us whether a TD type is more frequent over other TD types. Such findings help prioritizing measures like developing solutions to fix the more frequent TD types over the less frequent TD types. Trend analysis (RQ3) allows us to see how any specific TD type could increase (or decrease) quicker (or slower) than the others and whether a TD is on the rise in recent years. 
\subsection{Data Used for the Empirical Study} \label{sec:data_collection}
We collected all the 13.5K review comments between 2016 and 2021 from the two platforms, rOpenSci and BioConductor. As of 2021 and during the time of our analysis, rOpenSci has 173 and BioConductor as 1124 approved R packages. Our dataset also has $\sim$4.5K issue comments that are provided in response to the comments. We extracted these issue comments from respective peer-review GitHub repositories of each platforms \cite{github_ropensci,github_bioconductor} using web crawling and text parsing techniques. 
 Following Codabux et al. \cite{codabux2021technical}, we chose only the approved packages to ensure the completeness of the review process and the standard of the packages so that we can get a legitimate view of the TD from our study.  
\subsection{How prevalent are TDs in R packages? (RQ2)}\label{sec:rq2}
\subsubsection{Approach} 
 We automatically determined the TD types indicated in all the issue comments collected from rOpenSci and BioConductor packages. One issue comment can indicate one or more types of TD at different sentences or phrases. Since we are using sentence level detection, our developed ML models from RQ1 was able to catch all of the types expressed in each issue comment. We then counted and analyzed all the TD instances of different types both for rOpenSci and BioConductor packages.


\subsubsection{Results} 
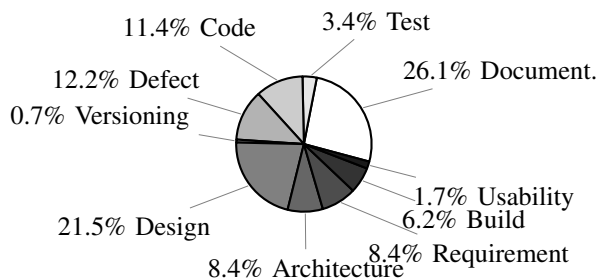
\begin{figure}[t]
	\centering

\begin{tikzpicture}[scale=0.3]

\pie[
        /tikz/every pin/.style={align=center},
        text=pin, number in legend,
        explode=0.0,
        rotate=-15,
        color={black!0, black!10, black!20, black!30, black!40, black!50, black!60, black!70, black!80, black!90}]
        {26.1/26.1\% Document.,
        3.4/3.4\% Test,
    11.4/11.4\% Code,
    12.2/12.2\% Defect,
    0.7/0.7\% Versioning,
    21.5/21.5\% Design,
    8.4/8.4\% Architecture,
    8.4/8.4\% Requirement,
    6.2/6.2\% Build,
    1.7/1.7\% Usability
    }
 
\end{tikzpicture}
	\caption{Overall distribution of different types of TD.}
	\vspace{-5mm}
	\label{fig:overall_td_distribution}
\end{figure}

We show the overall TD distribution of rOpenSci and BioConductor in combined in Figure \ref{fig:overall_td_distribution}. We see that documentation (26.1\%) and design (21.5\%) debts are two most prevalent TD in general. Table \ref{table_td_stats_ropensci_bio} depicts the distribution of different types of TD in rOpenSci and BioConductor individually in terms of frequency and percentage. It also shows some examples of respective TD that are predicted by our automated model. We find a total of 3938 TD instances in $\sim$4.5K issue comments of rOpenSci and 10159 TD instances in $\sim$9K issue comments of BioConductor. However, not all types of TDs are equally prevalent in R packages. Among all the types, documentation debt is the most frequent one with 938 occurrences ($\sim$23.8\%) in rOpenSci and 2740 occurrences ($\sim$27\%) in BioConductor. On the other side, versioning debt is the least prominent one with only 22 occurrences in rOpenSci and 71 occurrences in BioConductor. Apart from documentation debt, some other TD such as design debt, defect debt, and code debt occur frequently. Moreover, we see almost similar distribution of different TDs for both rOpenSci and BioConductor packages which suggests that different types of TDs occur in similar distribution in R packages regardless of the platform, scope, and field.   


\begin{table*}[t]
\caption{TD Distribution and Example Comment with TD in  rOpenSci and BioConductor}
\resizebox{\textwidth}{!}{
\begin{threeparttable}
\begin{tabular}{l|l|l} 
\toprule
{\textbf{TD Type}} & \bf{rOpenSci} & \bf{BioConductor}                                                                                                                                                        \\ 
\midrule
\multirow{2}{*}{\textbf{Documentation}}    & \TDstat{37}{938}{23.8}                                                                                                                                                                                                                                                       & \TDstat{109}{2740}{27}                                                                                                                                                                                                                  \\
                                  & \begin{tabular}[c]{@{}l@{}}\textit{``I think you should include a bit more info there: instructions for} \\ \textit{how to get it set up locally and a link to the nycflights13 package.'' \href{https://github.com/ropensci/software-review/issues/366}{\textit{ I\textsubscript{r366}}}\tnote{\textbf{*}}}\end{tabular}                                                                          & \begin{tabular}[c]{@{}l@{}}\textit{``In addition, your vignette must have code chunks primarily with eval = TRUE,} \\ \textit{so that the vignette accurately reflects the current capabilities of your software.''}\\\href{https://github.com/Bioconductor/Contributions/issues/585}{\textit{ I\textsubscript{b585}}\tnote{\textbf{$\dagger$}}}\end{tabular}
                                  
                                  \\
                                 
\midrule
\multirow{2}{*}{\textbf{Design}}           & \TDstat{29}{740}{18.8}                                                                                                                                                                                                                                                       & \TDstat{91}{2287}{22.5}                                                                                                                                                                                                                  \\
                                  &  
                                  
                                  \begin{tabular}[c]{@{}l@{}}\textit{``I think there are some opportunities to reduce dependencies in this package.''} \\ \href{https://github.com/ropensci/software-review/issues/274}{\textit{ I\textsubscript{r274}}}\end{tabular}
                                  
                                  & \begin{tabular}[c]{@{}l@{}}\textit{``These two functions are extremely similar and should be combined into one} \\ \textit{accounting for the minimal differences.'' \href{https://github.com/Bioconductor/Contributions/issues/1141}{\textit{ I\textsubscript{b1141}}}}\end{tabular}                                                          \\
\midrule
\multirow{2}{*}{\textbf{Defect}}           & \TDstat{20}{511}{13}                                                                                                                                                                                                                                                        & \TDstat{48}{1202}{11.8}                                                                                                                                                                                                                  \\
                                  & \begin{tabular}[c]{@{}l@{}}\textit{``There are cases where unreasonable values will be generated by the} \\ \textit{functions in this package (e.g., phone numbers with illegal area codes,}\\ \textit{weird names for people, etc.'' \href{https://github.com/ropensci/software-review/issues/94}{\textit{ I\textsubscript{r94}}}}\end{tabular}                                        & \begin{tabular}[c]{@{}l@{}}\textit{``I found the error is about the number of cores recruited by makeCluster()} \\ \textit{function.'' \href{https://github.com/Bioconductor/Contributions/issues/1943}{\textit{ I\textsubscript{b1943}}}}\end{tabular}                                                                                        \\

\midrule
\multirow{2}{*}{\textbf{Code}}             & \TDstat{17}{445}{11.3}                                                                                                                                                                                                                                                        & \TDstat{46}{1157}{11.4}                                                                                                                                                                                                                  \\
                                  & \begin{tabular}[c]{@{}l@{}}\textit{``I feel the user-facing function should be named ch\_gene\_sequence(),} \\ \textit{not ch\_sequence().'' \href{https://github.com/ropensci/software-review/issues/94}{\textit{ I\textsubscript{r94}}}}\end{tabular}                                                                                                                         & \begin{tabular}[c]{@{}l@{}}\textit{``Having arguments like reportPDF makes the function impure because the}\\ \textit{output depends on that argument.'' \href{https://github.com/Bioconductor/Contributions/issues/1941}{\textit{ I\textsubscript{b1941}}}}\end{tabular}                                                                      \\ 
\midrule
\multirow{2}{*}{\textbf{Requirement}}      & \TDstat{16}{406}{10.3}                                                                                                                                                                                                                                                        & \TDstat{31}{776}{7.6}                                                                                                                                                                                                                  \\
                                  & \begin{tabular}[c]{@{}l@{}}\textit{``It's not just dates you might want to subset, it could also be geographic}\\ \textit{space, or vertical levels within a climate model, or different file formats,}\\ \textit{or any number of other facets depending on the data set.'' \href{https://github.com/ropensci/software-review/issues/139}{\textit{ I\textsubscript{r139}}}}\end{tabular} & \begin{tabular}[c]{@{}l@{}}\textit{``Importing data from huge MAF files would again require a way to manipulate} \\ \textit{this kind of files from disk, as far as I know, such a feature is not availabe in} \\ \textit{maftools.'' \href{https://github.com/Bioconductor/Contributions/issues/1950}\textit{I\textsubscript{b1950}}}\end{tabular}  \\ 
\midrule
\multirow{2}{*}{\textbf{Architecture}}     & \TDstat{11}{293}{7.4}                                                                                                                                                                                                                                                        & \TDstat{35}{893}{8.8}                                                                                                                                                                                                                  \\
                                  & \begin{tabular}[c]{@{}l@{}}\textit{``A common solution is to split the general tool and the specific application} \\ \textit{into separate packages.''\href{https://github.com/ropensci/software-review/issues/102}{\textit{ I\textsubscript{r102}}}}\end{tabular}                                                                                                              & \begin{tabular}[c]{@{}l@{}}\textit{``Because of the size of the files we would recommend reformatting this package} \\ \textit{into and ExperimentHub package with the data stored on the Bioconductor} \\ \textit{AWS S3 buckets.'' \href{https://github.com/Bioconductor/Contributions/issues/498}\textit{I\textsubscript{b498}}}\end{tabular}   \\ 
\midrule
\multirow{2}{*}{\textbf{Build}}            & \TDstat{10}{259}{6.6}                                                                                                                                                                                                                                                        & \TDstat{25}{637}{6.3}                                                                                                                                                                                                                  \\
                                  & \textit{``Problems arose after installing using devtools::install\_all()'' \href{https://github.com/ropensci/software-review/issues/121}{\textit{ I\textsubscript{r121}}}}                                                                                                                                                                                              & \begin{tabular}[c]{@{}l@{}}\textit{``Looks like our Single Package Builder (SPB) failed to install WGSmapp} \\ \textit{before trying to run R CMD build on SCOPE.'' \href{https://github.com/Bioconductor/Contributions/issues/1242}\textit{I\textsubscript{b1242}}}\end{tabular}                                                           \\ 
\midrule
\multirow{2}{*}{\textbf{Test}}             & \TDstat{7}{199}{5.1}                                                                                                                                                                                                                                                        & \TDstat{12}{277}{2.7}                                                                                                                                                                                                                  \\
                                  & \begin{tabular}[c]{@{}l@{}}\textit{``A more robust test might also compare the values returned to the known}\\ \textit{tags from the REF website, or that the return type is in fact a tibble class}\\ \textit{object.'' \href{https://github.com/ropensci/software-review/issues/78}{\textit{ I\textsubscript{r78}}}}\end{tabular}                                                     & \begin{tabular}[c]{@{}l@{}}``\textit{There is an issue that there is no testing being done for the 'plotFemap'} \\ \textit{function.'' \href{https://github.com/Bioconductor/Contributions/issues/1897}\textit{I\textsubscript{b1897}}}\end{tabular}                                                                                        \\ 

\midrule
\multirow{2}{*}{\textbf{Usability}}        & \TDstat{5}{125}{3.2}                                                                                                                                                                                                                                                        & \TDstat{4}{119}{1.2}                                                                                                                                                                                                                  \\
                                  & \begin{tabular}[c]{@{}l@{}}\textit{``For the sake of your distracted future users, if you handled the auth}\\ \textit{failure more gracefully and gave the right helpful message, it would} \\ \textit{be a lot friendlier.'' \href{https://github.com/ropensci/software-review/issues/127}{\textit{ I\textsubscript{r127}}}}\end{tabular}                                                & \begin{tabular}[c]{@{}l@{}}\textit{``You should remove it as a system requirement so users don't install because} \\ \textit{they think its a requirement in order to use your package'' \href{https://github.com/Bioconductor/Contributions/issues/1897}\textit{I\textsubscript{b1897}}}\end{tabular}                                      \\ 
\midrule
\multirow{2}{*}{\textbf{Versioning}}       & \TDstat{1}{22}{.5}                                                                                                                                                                                                                                                        & \TDstat{3}{71}{.7}                                                                                                                                                                                                                      \\
                                  & \begin{tabular}[c]{@{}l@{}}\textit{``The version numbers for rgdal etc should be whatever the current version is,}\\ \textit{since that is what you are developing against.'' \href{https://github.com/ropensci/software-review/issues/22}{\textit{ I\textsubscript{r22}}}}\end{tabular}                                                                                       & \begin{tabular}[c]{@{}l@{}}\textit{``This is due to one of the core dependencies Seurat not installing properly due} \\ \textit{to their corrupted version 4.0.0.'' \href{https://github.com/Bioconductor/Contributions/issues/1845}\textit{I\textsubscript{b1845}}}\end{tabular}                                                           \\
\bottomrule
\end{tabular}%
\begin{tablenotes}\footnotesize
\item[*] Link to corresponding rOpenSci issue \textit{(I\textsubscript{rx})}
\item[$\dagger$] Link to corresponding BioConductor issue \textit{(I\textsubscript{bx})}
\end{tablenotes}
\end{threeparttable}
}
\label{table_td_stats_ropensci_bio}
\end{table*}






\subsection{How do the TDs evolve in R packages? (RQ3)}\label{sec:rq3}
\subsubsection{Approach}
\label{subsub:RQ3_Approach}
For each type and year, we divided the total number of TD instances detected in a year by the total number of packages in that year. Thus we determined the average number of TD (per package) over years and plotted them in the graph. Depending on the availability of data, we considered the packages from April 2015 to April 2020 for the trend analysis of rOpenSci and from April 2016 to April 2021 for BioConductor. To get an overall idea about the growth of different types of TD, we also calculated the overall change in the occurrence and the compound annual growth rate (CAGR) for each TD type combining the data of both rOpenSci and BioConductor over years 2016--2020.

\subsubsection{Results}
\begin{figure}[t]
  \centering
  \includegraphics[scale=.32]{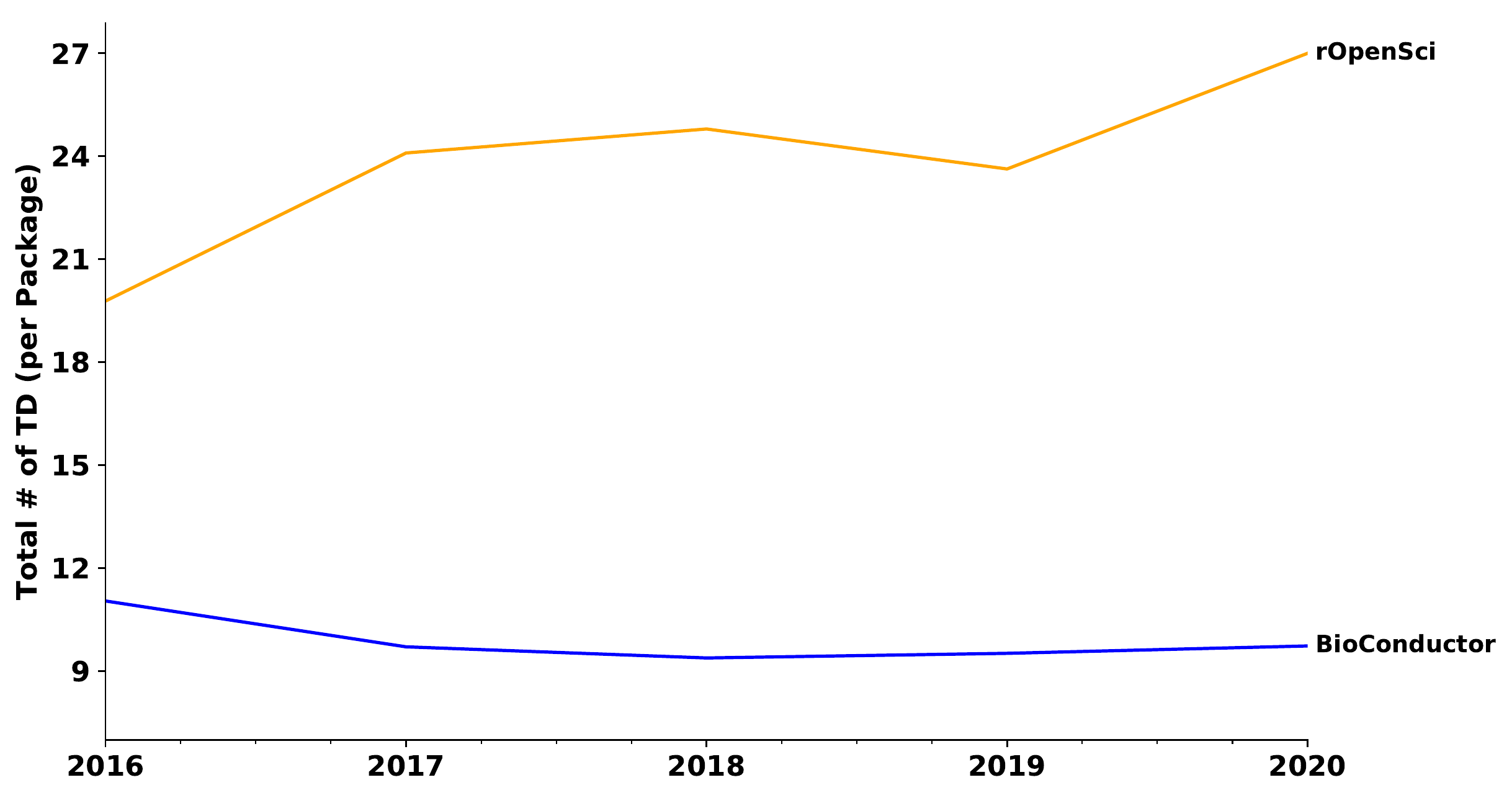}
  \caption{Evolution of TDs in rOpenSci and BioConductor}
  \label{fig:trend_Combined}
\vspace{-4mm}
\end{figure}

To see how TD differ among different types of platform and field of R, we compared our findings for rOpenSci (generic) and BioConductor (bioinformatics) packages. Though the distribution analysis of two platforms suggest that different types of TD occur in almost similar distribution regardless of the type and scope of R packages (see the result of RQ2), we observe significant differences in terms of quantity of TD and their evolution during our trend analysis. To get a more comprehensible view of these differences, we determined the total yearly number of TD instances (per package) aggregating all types both for rOpenSci and BioConductor, then plotted them on the same graph (Figure \ref{fig:trend_Combined}). We find that TD occur in significantly higher amount in generic platform i.e. rOpenSci compared to more domain-specific platform i.e. BioConductor. For example, from April 2016 to April 2020, the yearly average of per package TD instances for rOpenSci was around 24 while the number is less than 10 for BioConductor. Moreover, we see an overall upward trend of TD in rOpenSci packages with a steep slope at the end (April 2019 to April 2020) while the trend of BioConductor's TD is somewhat decreasing or stable. One possible reason might be unlike BioConductor, rOpenSci allows packages from diverse domains and scopes. Since different domains have different properties and vulnerabilities, rOpenSci packages are more prone to various types of TD.   


\begin{figure}[t]
  \centering
  \hspace*{-.1cm}%
   \includegraphics[scale=.36]{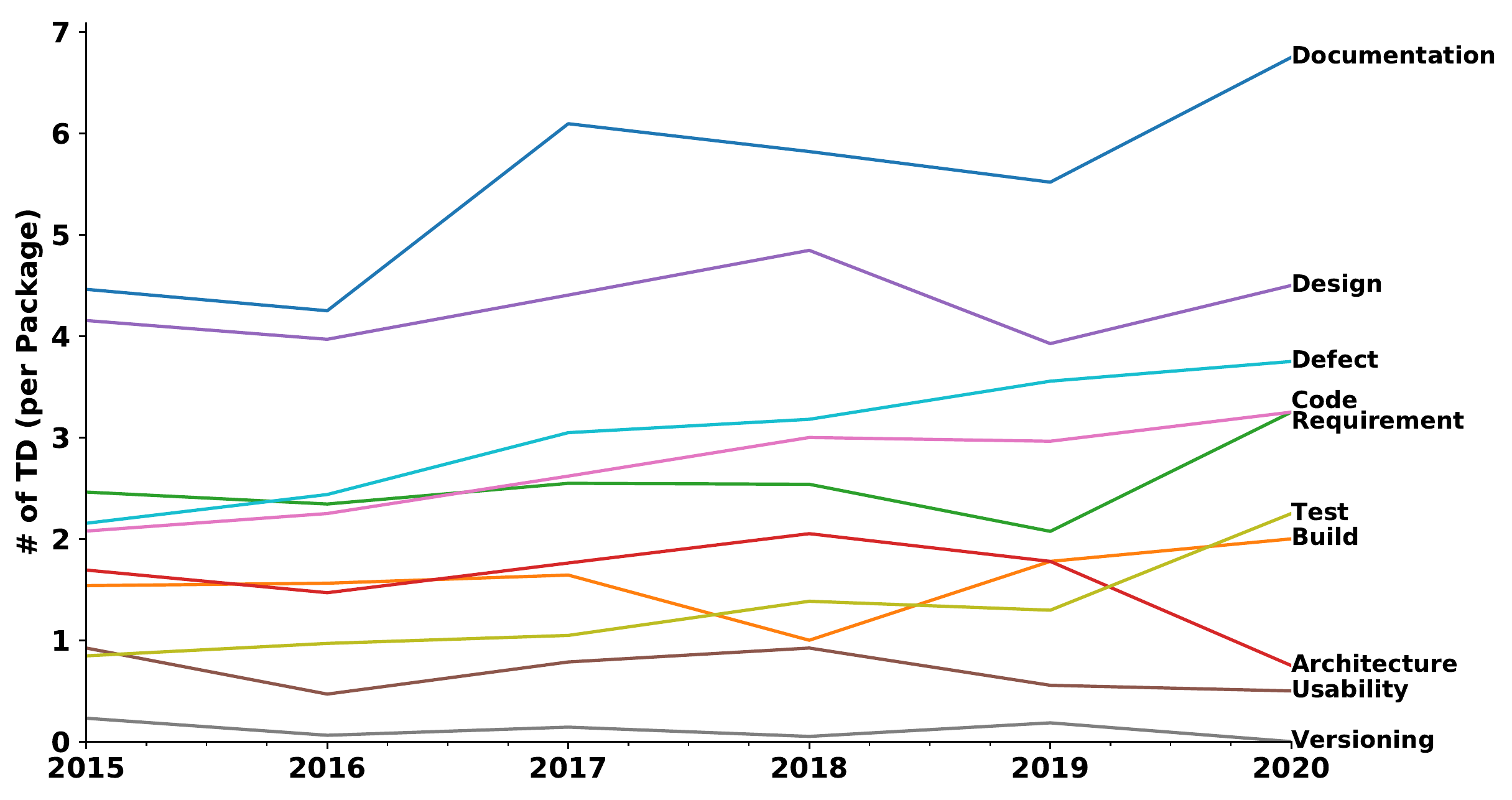}
  \caption{Trend analysis of different TD in rOpenSci}

  \label{fig:trend_rOpenSci}
\vspace{-4mm}
\end{figure}

\begin{figure}[t]
  \hspace*{-.1cm}%
   \includegraphics[scale=.36]{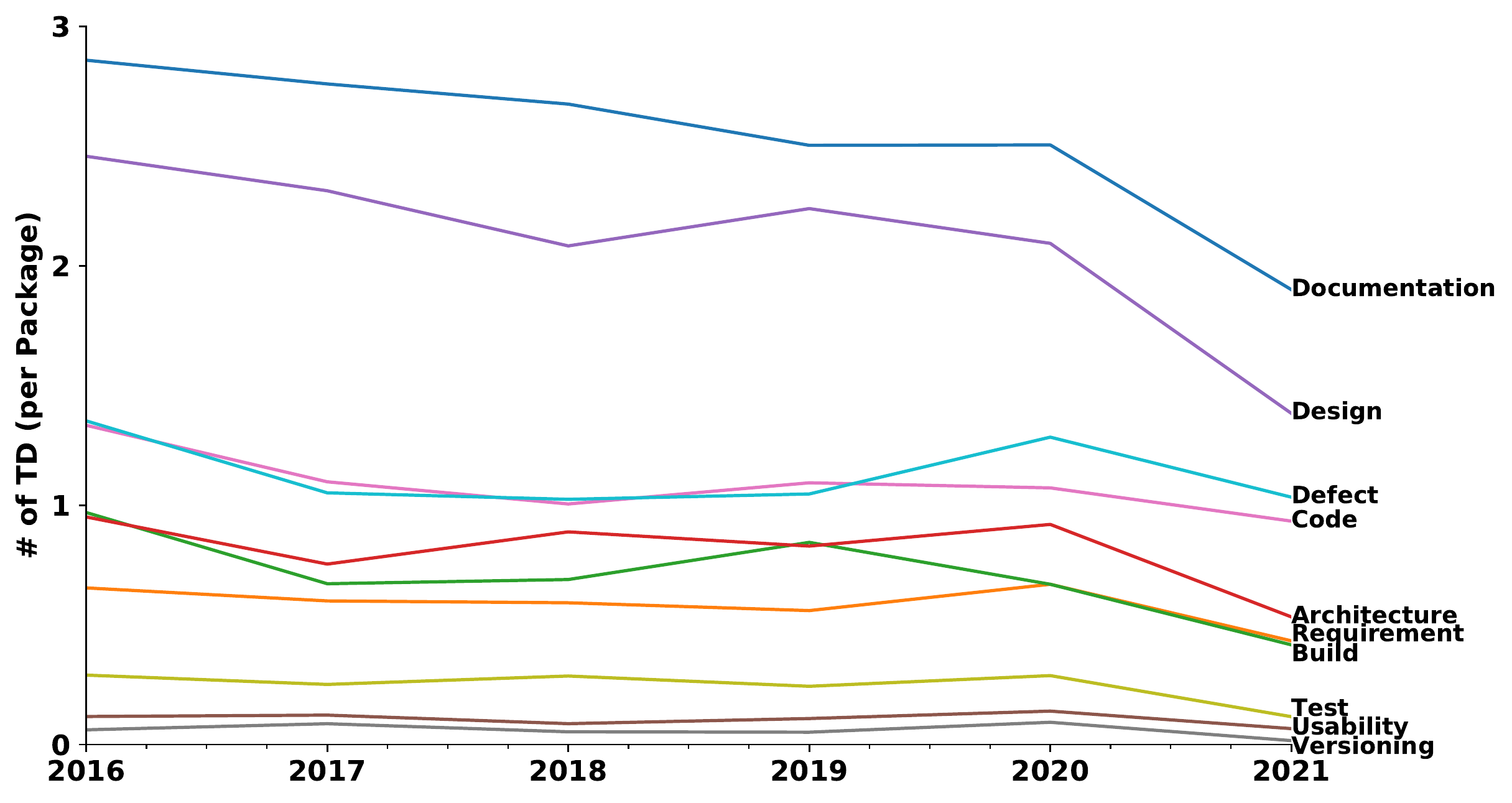}
  \caption{Trend analysis of different TD in BioConductor}

  \label{fig:trend_BioConductor}
\vspace{-4mm}
\end{figure}

Figure \ref{fig:trend_rOpenSci} and \ref{fig:trend_BioConductor} show the trend analysis of rOpenSci and BioConductor packages respectively. For rOpenSci, we see an overall upward trend for most types of TD (except the architecture debt). Among them, documentation debt has the most overall per package increment of $\sim$2.5 unit over five years starting at $\sim$4.5 debts in 2015 and reaching $\sim$7 in 2020 with a steep slope at the tail (Figure \ref{fig:trend_rOpenSci}). It indicates the constantly growing concern of the reviewers and users regarding documentation quality. Defect and test debts are two other types with significant increment of 1.6 and 1.4 unit per package throughout this time. On the other hand, we see a decreasing trend in architecture debt with less than 1 unit (per package) in 2020. Literature suggests that unlike other TD (such as code-level technical debt), architecture debt is very difficult to detect as they do not show any clear symptom \cite{verdecchia2020architectural, verdecchia2021building}. So it might happen that with growing diversity and complexity of packages, it is gradually becoming more challenging to identify architecture debts. For BioConductor, we see an overall downward trend for all types of TD. Even for the most prevalent type (i.e. documentation debt), the per package number has dropped below 2 in 2021 while in 2016 it was close to 3 (Figure \ref{fig:trend_BioConductor}). Table \ref{table:td_growth_rate} shows the overall growth-rate of the TDs. Documentation debt is the most growing TD among all with an annual increment of 2.14 units per package. On the other hand, test debt is the fastest one to grow with a growth-rate of 15.05\%. Other debts with significant increment and growth-rate are defect, code, build, and requirement debt.   

\begin{table}[t]
\centering
\caption{Growth-rate of different types of TD.}
\begin{tabular}{@{}lcr@{}}
\toprule
\textbf{TD Type} &  $\Delta$\textbf{Occurrence} & \textbf{Growth Rate (\%)} \\ \midrule
Documentation    & 2.14                & 5.41                      \\
Build            & 0.45                & 3.78                      \\
Requirement      & 0.61                & 3.42                      \\
Architecture     & -0.75               & -7.15                     \\
Design           & 0.17                & 0.52                      \\
Usability        & 0.05                & 1.77                      \\
Code             & 0.74                & 3.82                      \\
Versioning       & -0.03               & -5.58                     \\
Test             & 1.28                & 15.05                     \\
Defect           & 1.24                & 5.84                      \\ \bottomrule
\end{tabular}
\label{table:td_growth_rate}
\end{table}


\section{Discussions}\label{sec:discussion}
We first analyze whether the 10 TD types show any correlation in our studied empirical dataset (\sec\ref{sec:td-interdependence}). In \sec\ref{sec:comparison-zadia}, we compare the TD distribution between our empirical study findings with the state-of-the-art~\cite{codabux2021technical}. 

\subsection{Correlation between the 10 TD types}\label{sec:td-interdependence}
\begin{figure}[!htb]
  \centering
  \includegraphics[scale=.36]{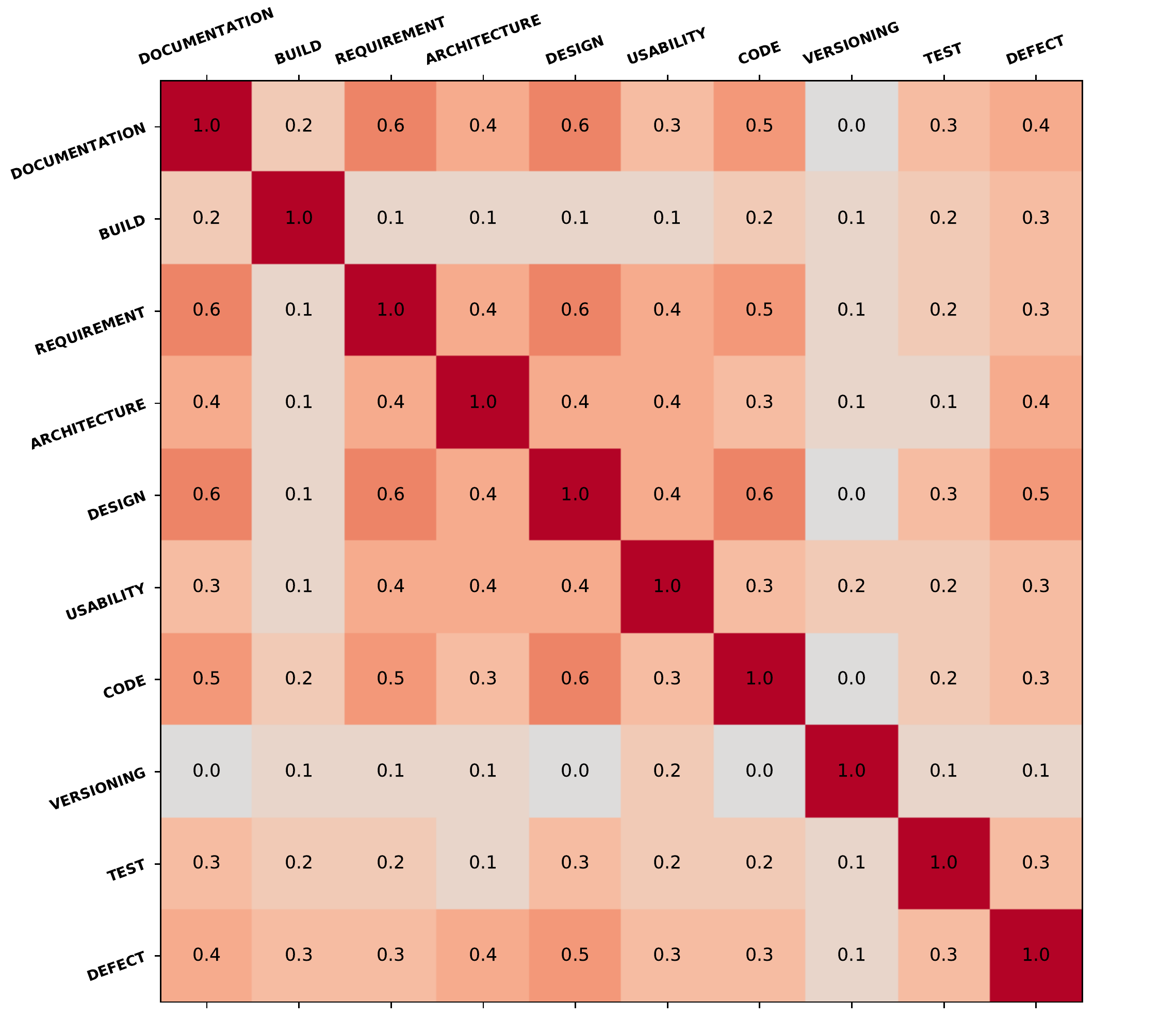}
  \caption{Correlation between different types of TD in R packages. Intensity of (red) color indicates the level of correlation.}

  \label{fig:TD_correlation}
\vspace{-5mm}
\end{figure}
Since a R package might contain multiple number of TD coming from different types, there is a possibility that these types are interdependent and correlated. For example, one type of TD might cause or contribute to another type of TD directly or indirectly. To investigate the hypothesis, we used Spearman's correlation coefficient to determine such correlation between different types of TD found in the entire dataset used for empirical study in \sec\ref{sec:data_collection}. Spearman's coefficient is a measure of monotonic relationship between two variables \cite{spearman_correlation_1}. It has a range of -1 to +1, with ±1 denoting perfect positive or negative correlation and 0 denoting no association. We report the Spearman's coefficients between each pair of TD types in Figure \ref{fig:TD_correlation}. While most of the technical debts are somewhat positively correlated, we observe strong positive correlations among 4 types (i.e. Documentation, Requirement, Design, and Code debts) with Spearman's coefficients of +0.5 to +0.6 indicating that they are highly likely to trigger each other at several cases. For example, design and code debts are similar in terms of pattern and origin i.e. both are related to coding style and quality \cite{code_design_relation_1,code_design_relation_2}. Hence, we see a strong relation between them. Documentation debt is closely associated with requirement debt and often caused by it as well \cite{documentation_requirement_relation, codabux2021technical}. Moreover, documentation greatly depends on code and design i.e. it is difficult to produce (and maintain) a good documentation for a code with bad programming style or poor design. Also, requirement debts are closely related to code and design debts since poor coding style or design often results in partial or incorrect implementation of requirements.

\subsection{Comparison with previous work}\label{sec:comparison-zadia}
\begin{table}
\centering
\caption{Comparison of TD distribution between Codabux et al. (\textbf{C}) \cite{codabux2021technical} and our study: \bf{R} = rOpenSci, \bf{O} = Overall.} 
\resizebox{\columnwidth}{!}
{
\begin{tabular}{ll} 
\toprule
\textbf{TD Type} & \textbf{Distribution Observed}                                                                                                            \\ 
\midrule
Documentation   & \threebars{31}{23.8}{26.1}                          \\ 
Design          &   \threebars{12.3}{18.8}{21.5}                \\ 
Defect         &   \threebars{12}{13}{12.2}                    \\ 
Code           &  \threebars{11.3}{14.5}{11.4}               \\ 
Requirement    &  \threebars{6.3}{10.3}{8.4} \\ 
Architecture   &   \threebars{5.8}{7.4}{8.4}  \\ 
Build          & \threebars{5.7}{6.6}{6.2} \\ 
Test           &    \threebars{7.5}{5.1}{3.4} \\ 
Usability      &   \threebars{4}{3.2}{1.7}   \\ 
Versioning      &    \threebars{0.8}{0.5}{0.7}         \\
 \midrule
 Platforms studied  & \textbf{C~}rOpenSci, We: rOpenSci \& ~BioConductor                                                                       \\ 
 \# Package Analyzed        & \textbf{\textbf{C }}157, We Studied: ~\textbf{R }173~~\textbf{Overall}~1,297                                  \\ 
 \# Issue comments  & \textbf{\textbf{C~}}458 We Studied: \textbf{R }4,500 \textbf{Overall} 13,500   \\
\bottomrule

\end{tabular}
}
\label{table:comparison_with_zadia}
\end{table}
Since Codabux et al. studied the distribution of different types of TDs in R packages of rOpenSci \cite{codabux2021technical}, we compared our finding with theirs. While Codabux et al. analyzed 600 review comments of 157 packages from rOpenSci platforms, we analyzed all the $\sim$13.5K review comments collected from all 1297 packages in two platforms, rOpenSci and BioConductor. 
In Table \ref{table:comparison_with_zadia}, we show the distribution of each of the 10 TD types in our dataset in two ways: in rOpenSci platform and overall (i.e., rOpenSci and BioConductor). We compare our distribution against those found by Codabux et al. in their studied rOpenSci packages. We see that our findings are mostly congruent to theirs with a few exceptions. While the most and the least prevalent types match in both studies (i.e. documentation and versioning debt), dissimilarity is observed in other cases. For example, according to Codabux et al. the second-most prevalent TD type is code debt (14.5\%) while according to our study it is design debt (18.8\% in rOpenSci and 21.5\% in overall). In summary, the overall percentages of different types of TDs differ between our study and Codabux et al. \cite{codabux2021technical} from moderate to large extents. Therefore, we complement Codabux et al. by offering a comprehensive overview of the prevalence of 10 TD types based on the excellent foundation laid out by Codabux et al. \cite{codabux2021technical}  (i.e., the benchmark dataset and the definition of 10 TD types).

\subsection{Threats to Validity}\label{sec:threats}
\it{Internal validity} threats relate to authors' bias 
while conducting the analysis. We mitigated the bias in our machine learning models 
by training, testing, and evaluating them using standard practices. There was no common data between the training and test set. We followed existing literature about TD and discussed among co-authors for listing different types of TD and detecting them.
\it{Construct validity} threats relate to the difficulty in finding
data for our study. Our data collection was exhaustive, as we processed more than 13.5K review comments from 1297 approved packages of two different popular platforms (i.e. rOpenSci, BioConductor).  
\it{External validity} threats relate to the
generalizability of our findings. We mitigated this threat by conducting our study on a large-scale data. We experimented with the peer review documentation of R packages to analyze TD. We used two different platforms and compared our findings with previous work as well. Moreover, the performance of our automated framework that we used for our empirical study was satisfactory. Hence, our findings are as generalized as possible and our finding can be applied to other similar platforms and applications. 
\section{Implications}\label{sec:implications}
Our TD detection tools and empirical study findings can be useful for the following stakeholders in SE: \begin{inparaenum}
\item \bf{R package maintainers} to continuously monitor the prevalence and evolution of TDs in their R packages, 
\item \bf{R package contributors} to stay aware of TDs in the open-source R packages and to decide to contribute to fix the TDs when needed,
\item \bf{R software developers} to decide which R packages to pick from among multiple competing choices,
\item \bf{R software engineering researchers} to develop techniques to automatically fix TDs to improve the adoption of the R packages by R practitioners, 
\item \bf{R package vendors and organizations} to make decisions on which R packages can be safer and/or more mature to be integrated into their enterprise R-based software toolkit, and
\item \bf{R educators} to educate and inform R practitioners of the TDs in R packages.
\end{inparaenum} We discuss the implications below.

\begin{figure}[t]
  \centering
  \includegraphics[scale=.47]{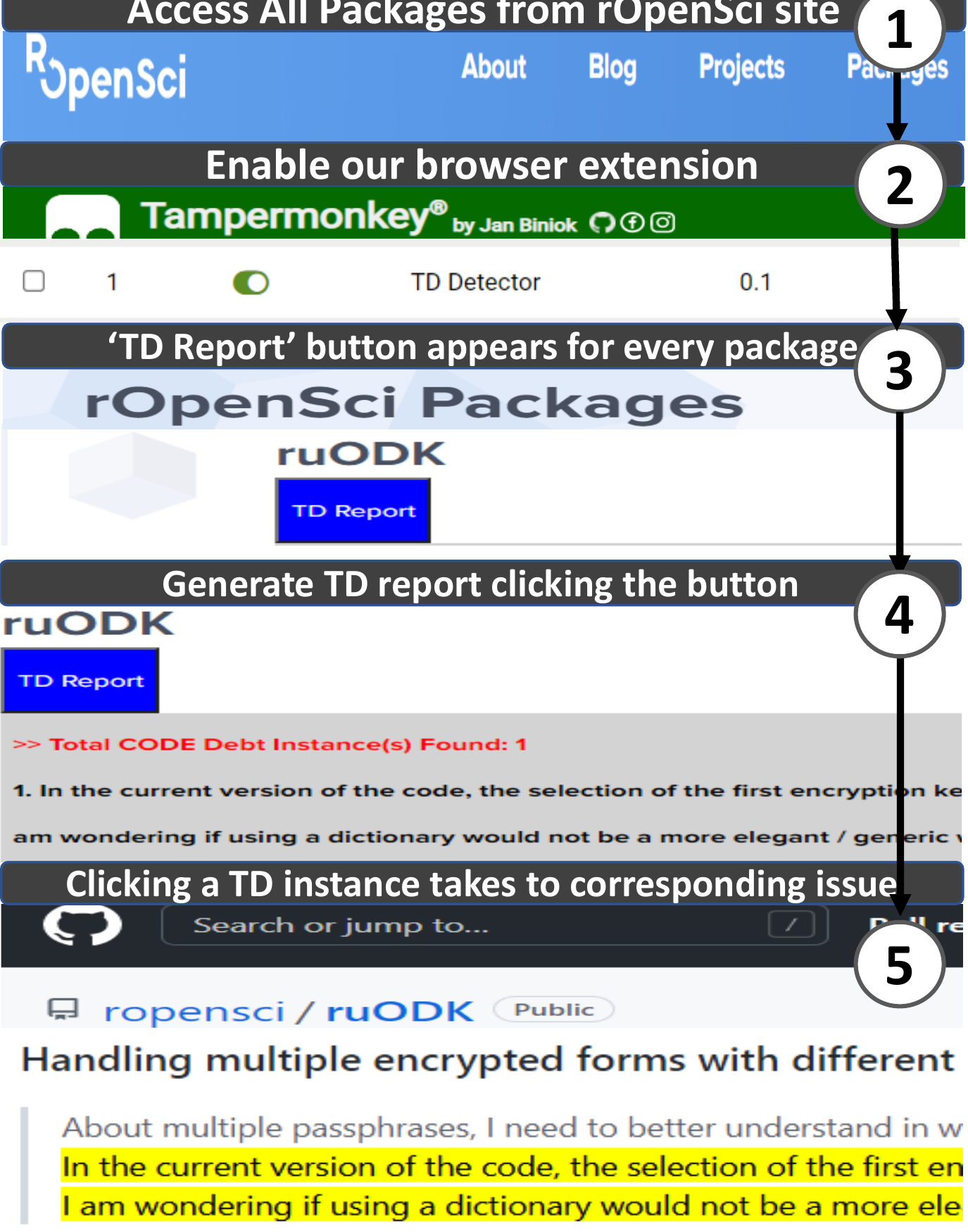}
  \caption{TD Alert Browser Plug-in in Existing Packages}

  \label{fig:browser_extension}
\vspace{-5mm}
\end{figure}

\bf{\ul{R Package Maintainers}} can use our TD detection models to monitor the presence of TDs in their packages. As we noted in \secs\ref{sec:introduction}, \ref{sec:related-work}, packages with TDs are more prone to issues like bugs and maintenance problems \cite{fernandez2014guiding, kruchten2012technical, li2015architectural, sultana2020examining, hall2014some}. Therefore, the presence of TDs can inform the package maintainers of deeper underlying issues with their packages, which then they can start analyzing to make proactive fixing instead of reactive fixing. To assist R package maintainers, we have developed a browser extension using TamperMonkey \cite{tamper_monkey} that will alert the users about any potential TD in a rOpenSci package using our automated framework. In Figure \ref{fig:browser_extension}, we show screenshots of our browser extension. First, we go to a website containing R packages (e.g., rOpenSci) \circled{1}. Second, we enable our browser extension \circled{2}. Third, a button (``TD Report'') will appear with every package available on the site \circled{3}. By clicking the button, we can see the detailed TD report (i.e. potential TD instances of different types) \circled{4}. We can also access the corresponding GitHub issue comment by clicking on a particular TD instance \circled{5}. The effectiveness analysis of the browser extension using R practitioners is our future work. 


\bf{\ul{R Package Contributors}} can utilize our browser extension (\fig\ref{fig:browser_extension}) to determine which package needs immediate help to reduce the number of TDs in it. Developer discussions in crowd forums GitHub and Stack Overflow about software libraries are found to be useful to fix bugs, enhance the software library, and so on~\cite{Uddin-OpinionValue-TSE2019}. Therefore, the TD reports produced by our browser extension can be useful to decide which modules of a R package needs more attention. For example, if an R package is showing a growing number of architecture debt, the contributor can collaborate with the corresponding package creator to redesign the package source code architecture. Indeed, we observe recommendations from R package contributors in our studied review comments to improve the architecture of an R package (\tbl\ref{table_td_stats_ropensci_bio}): \emt{A common solution is to split the general tool and the specific application into separate packages}. R package contributors can also help with testing the R packages, if they observed Test related TDs. For example, we found the following suggestion from a contributor in rOpenSci  (\tbl\ref{table_td_stats_ropensci_bio}): \emt{A more robust test might also compare the values returned to the known ....}. Indeed, design and architecture related TDs accounted for total 29.5\% of all TDs across all packages in rOpenSci and BioConductor (see \fig\ref{fig:overall_td_distribution}). Therefore, the detection and reporting of various TDs can help package maintainers and contributors work together to improve the quality of R packages.

\bf{\ul{R Software Developers}} are tasked with developing software products and a crucial component to any modern day software development is the use third-party software libraries like the open-source R packages need to stay aware of the potential problems and TDs in the R packages that they use, to ensure that they can make important decisions~\cite{Uddin-OpinionSurvey-TSE2019}. For example, research findings show that developers look for APIs/libraries with good documentation~\cite{Uddin-HowAPIDocumentationFails-IEEESW2015}, positive opinions about key aspects like usability of the package~\cite{Uddin-OpinerReviewAlgo-ASE2017,Uddin-OpinerReviewToolDemo-ASE2017,Uddin-APIAspectMining-TechReport2017}, and so on. From \fig\ref{fig:overall_td_distribution}, we can observe that the most prevalent TDs in the R packages (from rOpenSci and BioConductor) are related to lack of documentation (26\%). Therefore, R software developers need to be aware of sub-optimal documentation support in the R packages from rOpenSci and BioConductor, in genral. However, if a given functionality can be provided by multiple competing R packages provided in both rOpenSci and BioConductor platforms, we find from \tbl\ref{table_td_stats_ropensci_bio} that rOpenSci packages have on average less proportion (23.8\%) of documentation debt than those from BiConductor (27\%). Such information can be useful for a R software developer to determine which R package and platform to choose. 

\begin{figure}[t]
  \centering
   \includegraphics[scale=.41]{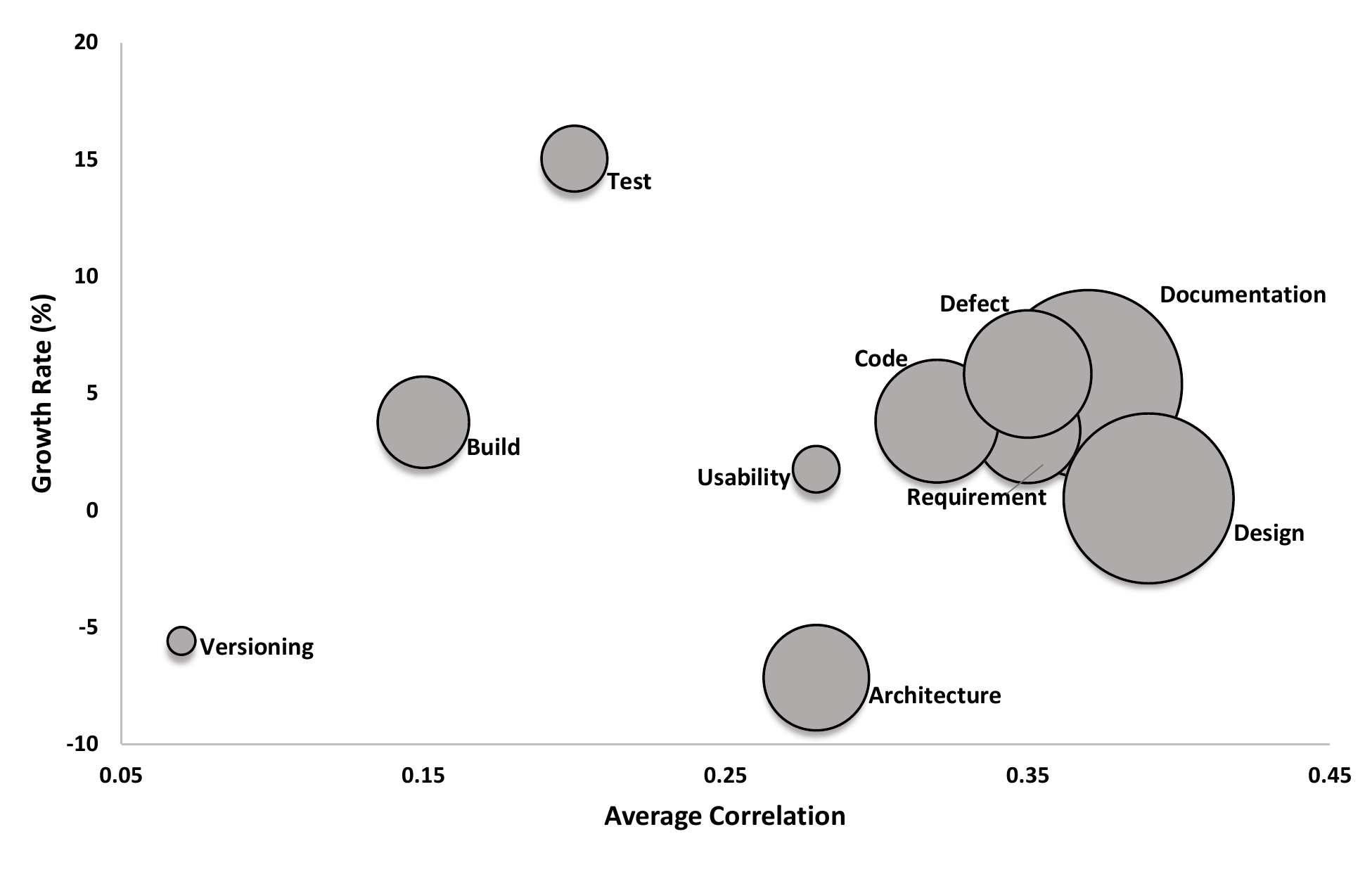}
  \caption{Perceived impact of the different types of TD. Circle size denotes \# of occurrences of corresponding TD}

  \label{fig:bubble_chart}
\vspace{-4mm}
\end{figure}
\bf{\ul{R Software Engineering (SE) Researchers}} can utilize the findings from our study to determine the major areas of focus for TDs in R packages and to develop tools and techniques to fix the TDs. As we noted in \fig\ref{fig:overall_td_distribution}, several TD types are more prevalent than others, e.g., documentation debts. Therefore, SE researchers can develop techniques to automatically improve the documentation debts in the R packages. This can be done by automatically creating new documentation, by enhancing the documentation with crowd-shared knowledge~\cite{Uddin-OpinerAPIUsageScenario-TOSEM2020,Uddin-OpinerAPIUsageScenario-IST2020,Chakraborty-NewLangSupportSO-IST2021}, or by automatically analyzing and fixing specific documentation smells in the package documentation~\cite{Khan-DocSmell-SANER2021}. During the fixing of the TDs, SE researchers can also benefit from the correlation between the TDs, i.e., when certain TD types seem to correlate with each other more than they correlate with other TDs. Intuitively, such correlation would denote that the correlated TD types may trigger each other. Indeed, in \fig\ref{fig:TD_correlation}, we showed that design and code debts correlate more with each other than they do with other debts (see \sec\ref{sec:discussion}). Intuitively, this means that if SE researchers can influence the development better code IDEs (with focus on module design) for R packages, the R package creators may hope to reduce the code and design debts in their packages.  The SE researchers can also benefit from the analysis of prevalence vs severity of the TDs in terms of how the TDs are observed empirically. To assist with this analysis, we further analyzed  the  impact  of  different types of TD  with respect to the frequency and the level of influence (i.e. trigger) it might cause on other TD (see  Figure \ref{fig:bubble_chart}).  We computed the growth-rate of each TD type (as described section \ref{subsub:RQ3_Approach}) and plotted them along x-axis. For example, the growth-rate of build debt is 3.78\%. For overall impact a type has towards other TD occurrences, we calculated the average correlation of each type and plotted them along y-axis. For example, we can compute the average correlation among documentation debt and all other 9 types from the first row of Figure \ref{fig:TD_correlation}) which is 0.37. The circle size indicates the total number of TD instances found in our empirical study for that type. From Figure \ref{fig:bubble_chart}, we observe that documentation debt is the most threatening type considering all the aspects together (i.e. size, influence, growth). The other types close to it in terms of overall impact are design, defect, code, and requirement debts. We also see that design debt has the most average correlation with others which indicates that it contributes the most to occurrence of other TD. Such analyses can inform SE researchers of the relative importance of fixing TDs like design debts (because these debts show high correlation with many debts).

\bf{\ul{R Package Vendors}} now account for almost all major cloud-based platform provider like Microsoft Azure as well as different enterprise data science toolkit creator (e.g., Alteryx). The growing adoption of ML-based solutions into diverse use cases of our daily life has only accelerated the needs for industrial-scale R packages. Therefore, R package vendors can use our browser extension to stay aware of the quality of open-source R packages and to decide which packages are safe to be included into their enterprise toolkit offerings. In the process, R package vendors can also collaborate with the open source R package creators to improve their packages, e.g., by creating test cases and by improving the code itself. Indeed, testing of R packages is important as we observed in \fig\ref{fig:bubble_chart}: though the total number and influence of test debt are not high, it is growing rapidly over the last few years with a growth rate of 15\% which summons more attention in the future.  

\bf{\ul{R Educators}} can develop tutorials and information sessions to educate R package creators and practitioners about the diverse TD types that can exist in their packages. Given that TDs can have long-term impact on the quality and bug-proneness of the underlying R packages, such education can help the R package creators to be mindful of the TDs and to take active actions to fix those. For example, during the requirement analysis of an R package, the developers can ensure the requirements are properly understood before being implemented. This is important given that requirement debts account for 8.4\% of all TDs in our study (\fig\ref{fig:overall_td_distribution}). In addition, the R educators can also develop targeted information session to show R package creators on how to document their code, to design their module properly, etc. The documentation and design debts accounted for total 47.6\% of all TDs (see \fig\ref{fig:overall_td_distribution}).

\section{Related Work} \label{sec:related-work}
Related works can broadly be divided into \bf{Studies} to understand and \bf{Techniques} to analyze and detect TDs.

\textbf{Studies.} The phrase `Technical debt' was first introduced in 1992 by Ward Cunningham \cite{cunningham1992wycash}. Existing literature defines TD as an implementation construct that is profitable in short term but problematic for future work and changes \cite{avgeriou2015reducing, allman2012managing,tom2013exploration, tom2012consolidated,li2015systematic}. Alves et al. devised a TD taxonomy with some new types i.e. people, process, service, usability debts \cite{alves2016identification}. Some other common types of TD discussed in various studies are social \cite{tamburri2013social}, data \cite{codabux2017empirical}, database \cite{al2016database}, infrastructure \cite{debois2008agile} debts. Behutiye et al. analyzed the concept of TD in the context of agile development \cite{behutiye2017analyzing}. Several researches also analyzed the impacts of TDs on the software development process. Studies have found that TDs are unavoidable and even desirable sometimes \cite{besker2018embracing}, but may lead to development crises in the long run \cite{martini2015investigating, kruchten2013technical, lim2012balancing}. TDs hinder the maintainability of software \cite{zazworka2011investigating, kruchten2012technical}, cause bugs and errors \cite{hall2014some}, create software vulnerabilities \cite{sultana2020examining}, and affect the productivity of software developers \cite{besker2018technical}. Hence, the repayment of TDs becomes very costly with time \cite{ampatzoglou2015financial, erdogmus1999comparative}. Unlike us, very few research, however, focused on TDs specific to R language except very recent effort by Codabux et al. \cite{codabux2021technical}. They investigated TD in the peer-review of rOpenSci R packages and developed a taxonomy of 10 types of TD (i.e. documentation) that persist in R packages, which we used in this paper.

\textbf{Techniques.} Different techniques or strategies have been proposed over the years to analyze and measure TDs, and manage them efficiently \cite{rios2018tertiary, alfayez2020systematic,nikolaidis2021experience}. One important step to report and manage different types of TD is to detect their existence automatically. Marinescu proposed a novel approach for assessing technical debt based on some metrics-based detection rules for design flaws i.e. violations of well-known design principles \cite{marinescu2012assessing}. Automated approaches like code smells detection, automated static analysis issues, and collection of code metrics were used to identify TD \cite{zazworka2013case, zazworka2014comparing}. Recently, Rantala explored natural language processing along with static code analysis to detect TD efficiently \cite{rantala2020towards}. Tsoukalas et al. evaluated the ability of machine learning methods to model and predict TD evolution in software application using source code \cite{tsoukalas2020technical}. Unlike the TDs we detect that are not explicitly labeled by a developer, SATD are debts that are intentionally committed and admitted by the programmers \cite{potdar2014exploratory}. Most of the recent works have focused on detection of self-admitted technical debts (SATD) \cite{da2017using, maldonado2017empirical,liu2018satd,wattanakriengkrai2018identifying,ren2019neural,yu2020identifying,xavier2020beyond}. In contrast, in this paper we focus on the detection and analysis of non-SATDs in R, i.e., TDs that are self-admitted, but nevertheless, are prevalent and thus should be studied. Unlike the existing papers, we focus specifically on R packages.
\section{Conclusions} \label{sec:conclusion}
R is a popular scientific programming language, with growing adoption in analytics, statistical and mission-critical software products. Therefore, the open source R packages need to be studied for quality assurance like the presence of Technical Debts (TDs). We developed ML models to detect TD instances in existing packages from peer-review issue comments. We then automatically labeled a large number of R package review comments with our automated framework and used them for our empirical study. Our distribution and trend analyses show that TD frequency and growth-rate in R packages varies from type to type. Our correlation analysis provides strong evidence of interdependence among several TD types. Further analysis reveals that documentation debt is the most impactful one in terms of prevalence, growth, and influence. We developed a browser extension using our automated framework that alerts the users about TD instances in existing rOpenSci packages. Our automated framework and empirical study findings can be leveraged for similar study involving diverse language and platforms. It can also be used to detect TD in various existing packages using the issue comments of their respective repositories.

\section*{Acknowledgments}
This research work was funded by grants and awards from Natural Sciences and Engineering Research Council of Canada (NSERC), University of Calgary, Alberta Innovates, and Alberta Graduate Excellence Scholarship.

\begin{small}
\bibliographystyle{abbrv}
\bibliography{consolidated}

\begin{thebibliography}{10}

\bibitem{github_bioconductor}
Bioconductor.
\newblock \url{https://github.com/Bioconductor}.
\newblock Accessed: 2021-10-20.

\bibitem{github_ropensci}
ropensci software peer review.
\newblock \url{https://github.com/ropensci/software-review}.
\newblock Accessed: 2021-10-20.

\bibitem{tamper_monkey}
Tamper monkey.
\newblock \url{https://www.tampermonkey.net/}.
\newblock Accessed: 2021-10-20.

\bibitem{al2016database}
M.~Al-Barak and R.~Bahsoon.
\newblock Database design debts through examining schema evolution.
\newblock In {\em 2016 IEEE 8th International Workshop on Managing Technical
  Debt (MTD)}, pages 17--23. IEEE, 2016.

\bibitem{alfayez2020systematic}
R.~Alfayez, W.~Alwehaibi, R.~Winn, E.~Venson, and B.~Boehm.
\newblock A systematic literature review of technical debt prioritization.
\newblock In {\em Proceedings of the 3rd International Conference on Technical
  Debt}, pages 1--10, 2020.

\bibitem{allman2012managing}
E.~Allman.
\newblock Managing technical debt.
\newblock {\em Communications of the ACM}, 55(5):50--55, 2012.

\bibitem{alves2016identification}
N.~S. Alves, T.~S. Mendes, M.~G. de~Mendon{\c{c}}a, R.~O. Sp{\'\i}nola,
  F.~Shull, and C.~Seaman.
\newblock Identification and management of technical debt: A systematic mapping
  study.
\newblock {\em Information and Software Technology}, 70:100--121, 2016.

\bibitem{ampatzoglou2015financial}
A.~Ampatzoglou, A.~Ampatzoglou, A.~Chatzigeorgiou, and P.~Avgeriou.
\newblock The financial aspect of managing technical debt: A systematic
  literature review.
\newblock {\em Information and Software Technology}, 64:52--73, 2015.

\bibitem{arabie1996hierarchical}
P.~Arabie, L.~Hubert, G.~De~Soete, and A.~Gordon.
\newblock Hierarchical classification.
\newblock {\em ArabieP., HubertL., De SoeteG., \& GordonA., Clustering and
  classification}, pages 65--121, 1996.

\bibitem{avgeriou2015reducing}
P.~Avgeriou, P.~Kruchten, R.~L. Nord, I.~Ozkaya, and C.~Seaman.
\newblock Reducing friction in software development.
\newblock {\em Ieee software}, 33(1):66--73, 2015.

\bibitem{avgeriou2016managing}
P.~Avgeriou, P.~Kruchten, I.~Ozkaya, and C.~Seaman.
\newblock Managing technical debt in software engineering (dagstuhl seminar
  16162).
\newblock In {\em Dagstuhl Reports}, volume~6. Schloss Dagstuhl-Leibniz-Zentrum
  fuer Informatik, 2016.

\bibitem{behutiye2017analyzing}
W.~N. Behutiye, P.~Rodr{\'\i}guez, M.~Oivo, and A.~Tosun.
\newblock Analyzing the concept of technical debt in the context of agile
  software development: A systematic literature review.
\newblock {\em Information and Software Technology}, 82:139--158, 2017.

\bibitem{bellomo2016got}
S.~Bellomo, R.~L. Nord, I.~Ozkaya, and M.~Popeck.
\newblock Got technical debt? surfacing elusive technical debt in issue
  trackers.
\newblock In {\em 2016 IEEE/ACM 13th Working Conference on Mining Software
  Repositories (MSR)}, pages 327--338. IEEE, 2016.

\bibitem{besker2018technical}
T.~Besker, A.~Martini, and J.~Bosch.
\newblock Technical debt cripples software developer productivity: a
  longitudinal study on developers' daily software development work.
\newblock In {\em Proceedings of the 2018 International Conference on Technical
  Debt}, pages 105--114, 2018.

\bibitem{besker2018embracing}
T.~Besker, A.~Martini, R.~E. Lokuge, K.~Blincoe, and J.~Bosch.
\newblock Embracing technical debt, from a startup company perspective.
\newblock In {\em 2018 IEEE International Conference on Software Maintenance
  and Evolution (ICSME)}, pages 415--425. IEEE, 2018.

\bibitem{smote_prob_for_text}
R.~Blagus and L.~Lusa.
\newblock Smote for high-dimensional class-imbalanced data.
\newblock {\em BMC bioinformatics}, 14:106, 03 2013.

\bibitem{boettiger2015building}
C.~Boettiger, S.~Chamberlain, E.~Hart, and K.~Ram.
\newblock Building software, building community: lessons from the ropensci
  project.
\newblock {\em Journal of open research software}, 3(1), 2015.

\bibitem{code_design_relation_1}
F.~Buschmann.
\newblock To pay or not to pay technical debt.
\newblock {\em IEEE software}, 28(6):29--31, 2011.

\bibitem{chaitra2018review}
P.~Chaitra and D.~R.~S. Kumar.
\newblock A review of multi-class classification algorithms.
\newblock {\em Int. J. Pure Appl. Math}, 118(14):17--26, 2018.

\bibitem{Chakraborty-NewLangSupportSO-IST2021}
P.~Chakraborty, R.~Shahriyar, A.~Iqbal, and G.~Uddin.
\newblock How do developers discuss and support new programming languages in
  technical q\&a site? an empirical study of go, swift, and rust in stack
  overflow.
\newblock {\em Information and Software Technology (IST)}, page~19, 2021.

\bibitem{codabux2021technical}
Z.~Codabux, M.~Vidoni, and F.~H. Fard.
\newblock Technical debt in the peer-review documentation of r packages: a
  ropensci case study.
\newblock {\em arXiv preprint arXiv:2103.09340}, 2021.

\bibitem{codabux2017empirical}
Z.~Codabux, B.~J. Williams, G.~L. Bradshaw, and M.~Cantor.
\newblock An empirical assessment of technical debt practices in industry.
\newblock {\em Journal of software: Evolution and Process}, 29(10):e1894, 2017.

\bibitem{colas2006comparison}
F.~Colas and P.~Brazdil.
\newblock Comparison of svm and some older classification algorithms in text
  classification tasks.
\newblock In {\em IFIP International Conference on Artificial Intelligence in
  Theory and Practice}, pages 169--178. Springer, 2006.

\bibitem{cunningham1992wycash}
W.~Cunningham.
\newblock The wycash portfolio management system.
\newblock {\em ACM SIGPLAN OOPS Messenger}, 4(2):29--30, 1992.

\bibitem{da2017using}
E.~da~Silva~Maldonado, E.~Shihab, and N.~Tsantalis.
\newblock Using natural language processing to automatically detect
  self-admitted technical debt.
\newblock {\em IEEE Transactions on Software Engineering}, 43(11):1044--1062,
  2017.

\bibitem{debois2008agile}
P.~Debois.
\newblock Agile infrastructure and operations: how infra-gile are you?
\newblock In {\em Agile 2008 Conference}, pages 202--207. IEEE, 2008.

\bibitem{Delvin-BERTArch-Arxiv2018}
J.~Devlin, M.-W. Chang, K.~Lee, and K.~Toutanova.
\newblock {BERT}: Pre-training of deep bidirectional transformers for language
  understanding.
\newblock Technical report, \url{https://arxiv.org/abs/1810.04805}, 2018.

\bibitem{dumais1998using}
S.~Dumais et~al.
\newblock Using svms for text categorization.
\newblock {\em IEEE Intelligent Systems}, 13(4):21--23, 1998.

\bibitem{erdogmus1999comparative}
H.~Erdogmus.
\newblock Comparative evaluation of software development strategies based on
  net present value.
\newblock In {\em International Workshop on Economics-Driven Software
  Engineering Research EDSER}, volume~1, 1999.

\bibitem{fernandez2014guiding}
C.~Fern{\'a}ndez-S{\'a}nchez, J.~D{\'\i}az, J.~P{\'e}rez, and J.~Garbajosa.
\newblock Guiding flexibility investment in agile architecting.
\newblock In {\em 2014 47th Hawaii International Conference on System
  Sciences}, pages 4807--4816. IEEE, 2014.

\bibitem{gentleman2004bioconductor}
R.~C. Gentleman, V.~J. Carey, D.~M. Bates, B.~Bolstad, M.~Dettling, S.~Dudoit,
  B.~Ellis, L.~Gautier, Y.~Ge, J.~Gentry, et~al.
\newblock Bioconductor: open software development for computational biology and
  bioinformatics.
\newblock {\em Genome biology}, 5(10):1--16, 2004.

\bibitem{bert_success_2}
S.~Gonz{\'a}lez-Carvajal and E.~C. Garrido-Merch{\'a}n.
\newblock Comparing bert against traditional machine learning text
  classification.
\newblock {\em arXiv preprint arXiv:2005.13012}, 2020.

\bibitem{graves2005framewise}
A.~Graves and J.~Schmidhuber.
\newblock Framewise phoneme classification with bidirectional lstm and other
  neural network architectures.
\newblock {\em Neural networks}, 18(5-6):602--610, 2005.

\bibitem{hall2014some}
T.~Hall, M.~Zhang, D.~Bowes, and Y.~Sun.
\newblock Some code smells have a significant but small effect on faults.
\newblock {\em ACM Transactions on Software Engineering and Methodology
  (TOSEM)}, 23(4):1--39, 2014.

\bibitem{Khan-DocSmell-SANER2021}
J.~Y. Khan, M.~T.~I. Khondaker, G.~Uddin, and A.~Iqbal.
\newblock Automatic detection of five api documentation smells:
  Practitioners’ perspectives.
\newblock In {\em IEEE International Conference on Software Analysis, Evolution
  and Reengineering (SANER)}, page~12, 2021.

\bibitem{kingma2014adam}
D.~P. Kingma and J.~Ba.
\newblock Adam: A method for stochastic optimization.
\newblock {\em arXiv preprint arXiv:1412.6980}, 2014.

\bibitem{kruchten2012technical}
P.~Kruchten, R.~L. Nord, and I.~Ozkaya.
\newblock Technical debt: From metaphor to theory and practice.
\newblock {\em Ieee software}, 29(6):18--21, 2012.

\bibitem{kruchten2013technical}
P.~Kruchten, R.~L. Nord, I.~Ozkaya, and D.~Falessi.
\newblock Technical debt: towards a crisper definition report on the 4th
  international workshop on managing technical debt.
\newblock {\em ACM SIGSOFT Software Engineering Notes}, 38(5):51--54, 2013.

\bibitem{bert_success_5}
X.~Li, L.~Bing, W.~Zhang, and W.~Lam.
\newblock Exploiting bert for end-to-end aspect-based sentiment analysis.
\newblock {\em arXiv preprint arXiv:1910.00883}, 2019.

\bibitem{li2015systematic}
Z.~Li, P.~Avgeriou, and P.~Liang.
\newblock A systematic mapping study on technical debt and its management.
\newblock {\em Journal of Systems and Software}, 101:193--220, 2015.

\bibitem{li2015architectural}
Z.~Li, P.~Liang, and P.~Avgeriou.
\newblock Architectural technical debt identification based on architecture
  decisions and change scenarios.
\newblock In {\em 2015 12th Working IEEE/IFIP Conference on Software
  Architecture}, pages 65--74. IEEE, 2015.

\bibitem{lim2012balancing}
E.~Lim, N.~Taksande, and C.~Seaman.
\newblock A balancing act: What software practitioners have to say about
  technical debt.
\newblock {\em IEEE software}, 29(6):22--27, 2012.

\bibitem{synonym_replace_text_aug_3}
P.~Liu, X.~Wang, C.~Xiang, and W.~Meng.
\newblock A survey of text data augmentation.
\newblock In {\em 2020 International Conference on Computer Communication and
  Network Security (CCNS)}, pages 191--195. IEEE, 2020.

\bibitem{liu2018satd}
Z.~Liu, Q.~Huang, X.~Xia, E.~Shihab, D.~Lo, and S.~Li.
\newblock Satd detector: A text-mining-based self-admitted technical debt
  detection tool.
\newblock In {\em Proceedings of the 40th International Conference on Software
  Engineering: Companion Proceeedings}, pages 9--12, 2018.

\bibitem{adam_w}
I.~Loshchilov and F.~Hutter.
\newblock Decoupled weight decay regularization.
\newblock {\em arXiv preprint arXiv:1711.05101}, 2017.

\bibitem{maldonado2017empirical}
E.~d.~S. Maldonado, R.~Abdalkareem, E.~Shihab, and A.~Serebrenik.
\newblock An empirical study on the removal of self-admitted technical debt.
\newblock In {\em 2017 IEEE International Conference on Software Maintenance
  and Evolution (ICSME)}, pages 238--248. IEEE, 2017.

\bibitem{Manning-IRIntroBook-Cambridge2009}
C.~D. Manning, P.~Raghavan, and H.~Sch\"{u}tze.
\newblock {\em An Introduction to Information Retrieval}.
\newblock Cambridge Uni Press, 2009.

\bibitem{marinescu2012assessing}
R.~Marinescu.
\newblock Assessing technical debt by identifying design flaws in software
  systems.
\newblock {\em IBM Journal of Research and Development}, 56(5):9--1, 2012.

\bibitem{martini2015investigating}
A.~Martini, J.~Bosch, and M.~Chaudron.
\newblock Investigating architectural technical debt accumulation and
  refactoring over time: A multiple-case study.
\newblock {\em Information and Software Technology}, 67:237--253, 2015.

\bibitem{miller1995wordnet}
G.~A. Miller.
\newblock Wordnet: a lexical database for english.
\newblock {\em Communications of the ACM}, 38(11):39--41, 1995.

\bibitem{spectralCluster1}
A.~Y. Ng, M.~I. Jordan, and Y.~Weiss.
\newblock On spectral clustering: Analysis and an algorithm.
\newblock In {\em Advances in neural information processing systems}, pages
  849--856, 2002.

\bibitem{nikolaidis2021experience}
N.~Nikolaidis, D.~Zisis, A.~Ampatzoglou, A.~Chatzigeorgiou, and D.~Soudris.
\newblock Experience with managing technical debt in scientific software
  development using the exa2pro framework.
\newblock {\em IEEE Access}, 9:72524--72534, 2021.

\bibitem{pennington2014glove}
J.~Pennington, R.~Socher, and C.~D. Manning.
\newblock Glove: Global vectors for word representation.
\newblock In {\em Proceedings of the 2014 conference on empirical methods in
  natural language processing (EMNLP)}, pages 1532--1543, 2014.

\bibitem{potdar2014exploratory}
A.~Potdar and E.~Shihab.
\newblock An exploratory study on self-admitted technical debt.
\newblock In {\em 2014 IEEE International Conference on Software Maintenance
  and Evolution}, pages 91--100. IEEE, 2014.

\bibitem{early_stop_bert_2}
L.~Prechelt.
\newblock Automatic early stopping using cross validation: quantifying the
  criteria.
\newblock {\em Neural Networks}, 11(4):761--767, 1998.

\bibitem{early_stop_bert_1}
L.~Prechelt.
\newblock Early stopping-but when?
\newblock In {\em Neural Networks: Tricks of the trade}, pages 55--69.
  Springer, 1998.

\bibitem{rantala2020towards}
L.~Rantala.
\newblock Towards better technical debt detection with nlp and machine learning
  methods.
\newblock In {\em 2020 IEEE/ACM 42nd International Conference on Software
  Engineering: Companion Proceedings (ICSE-Companion)}, pages 242--245. IEEE,
  2020.

\bibitem{ren2019neural}
X.~Ren, Z.~Xing, X.~Xia, D.~Lo, X.~Wang, and J.~Grundy.
\newblock Neural network-based detection of self-admitted technical debt: From
  performance to explainability.
\newblock {\em ACM transactions on software engineering and methodology
  (TOSEM)}, 28(3):1--45, 2019.

\bibitem{rios2018tertiary}
N.~Rios, M.~G. de~Mendon{\c{c}}a~Neto, and R.~O. Sp{\'\i}nola.
\newblock A tertiary study on technical debt: Types, management strategies,
  research trends, and base information for practitioners.
\newblock {\em Information and Software Technology}, 102:117--145, 2018.

\bibitem{documentation_requirement_relation}
N.~Rios, L.~Mendes, C.~Cerdeiral, A.~P.~F. Magalh{\~a}es, B.~Perez, D.~Correal,
  H.~Astudillo, C.~Seaman, C.~Izurieta, G.~Santos, et~al.
\newblock Hearing the voice of software practitioners on causes, effects, and
  practices to deal with documentation debt.
\newblock In {\em International Working Conference on Requirements Engineering:
  Foundation for Software Quality}, pages 55--70. Springer, 2020.

\bibitem{binary_cross_AreLossFunctionSame}
L.~Rosasco, E.~D. Vito, A.~Caponnetto, M.~Piana, and A.~Verri.
\newblock Are loss functions all the same?
\newblock {\em Neural Computation}, 16(5):1063--1076, 2004.

\bibitem{seaman2011measuring}
C.~Seaman and Y.~Guo.
\newblock Measuring and monitoring technical debt.
\newblock In {\em Advances in Computers}, volume~82, pages 25--46. Elsevier,
  2011.

\bibitem{bag_of_words_for_text_1}
F.~Sebastiani.
\newblock Machine learning in automated text categorization.
\newblock {\em ACM computing surveys (CSUR)}, 34(1):1--47, 2002.

\bibitem{sierra2019survey}
G.~Sierra, E.~Shihab, and Y.~Kamei.
\newblock A survey of self-admitted technical debt.
\newblock {\em Journal of Systems and Software}, 152:70--82, 2019.

\bibitem{silla2011survey}
C.~N. Silla and A.~A. Freitas.
\newblock A survey of hierarchical classification across different application
  domains.
\newblock {\em Data Mining and Knowledge Discovery}, 22(1):31--72, 2011.

\bibitem{confusionMatrix_for_hierarchy}
D.~Silva-Palacios, C.~Ferri, and M.~J. Ram{\'\i}rez-Quintana.
\newblock Improving performance of multiclass classification by inducing class
  hierarchies.
\newblock {\em Procedia Computer Science}, 108:1692--1701, 2017.

\bibitem{spearman_correlation_1}
C.~Spearman.
\newblock The proof and measurement of association between two things.
\newblock 1961.

\bibitem{sultana2020examining}
K.~Z. Sultana, Z.~Codabux, and B.~Williams.
\newblock Examining the relationship of code and architectural smells with
  software vulnerabilities.
\newblock In {\em 2020 27th Asia-Pacific Software Engineering Conference
  (APSEC)}, pages 31--40. IEEE, 2020.

\bibitem{bert_fine_tuning}
C.~Sun, X.~Qiu, Y.~Xu, and X.~Huang.
\newblock How to fine-tune bert for text classification?
\newblock In {\em China National Conference on Chinese Computational
  Linguistics}, pages 194--206. Springer, 2019.

\bibitem{tamburri2013social}
D.~A. Tamburri, P.~Kruchten, P.~Lago, and H.~van Vliet.
\newblock What is social debt in software engineering?
\newblock In {\em 2013 6th International Workshop on Cooperative and Human
  Aspects of Software Engineering (CHASE)}, pages 93--96. IEEE, 2013.

\bibitem{bert_success_1}
I.~Tenney, D.~Das, and E.~Pavlick.
\newblock Bert rediscovers the classical nlp pipeline.
\newblock {\em arXiv preprint arXiv:1905.05950}, 2019.

\bibitem{tom2012consolidated}
E.~Tom, A.~Aurum, and R.~Vidgen.
\newblock A consolidated understanding of technical debt.
\newblock 2012.

\bibitem{tom2013exploration}
E.~Tom, A.~Aurum, and R.~Vidgen.
\newblock An exploration of technical debt.
\newblock {\em Journal of Systems and Software}, 86(6):1498--1516, 2013.

\bibitem{tsoukalas2020technical}
D.~Tsoukalas, D.~Kehagias, M.~Siavvas, and A.~Chatzigeorgiou.
\newblock Technical debt forecasting: an empirical study on open-source
  repositories.
\newblock {\em Journal of Systems and Software}, 170:110777, 2020.

\bibitem{Uddin-OpinionSurvey-TSE2019}
G.~Uddin, O.~Baysal, L.~Guerroj, and F.~Khomh.
\newblock Understanding how and why developers seek and analyze api related
  opinions.
\newblock {\em IEEE Transactions on Software Engineering}, page~40, 2019.

\bibitem{Uddin-OpinerReviewAlgo-ASE2017}
G.~Uddin and F.~Khomh.
\newblock Automatic summarization of {API} reviews.
\newblock In {\em Proc. 32nd IEEE/ACM International Conference on Automated
  Software Engineering}, page~12, 2017.

\bibitem{Uddin-APIAspectMining-TechReport2017}
G.~Uddin and F.~Khomh.
\newblock Mining api aspects in api reviews.
\newblock Technical report,
  \url{https://swat.polymtl.ca/data/opinionvalue-technical-report.pdf}, 2017.

\bibitem{Uddin-OpinerReviewToolDemo-ASE2017}
G.~Uddin and F.~Khomh.
\newblock Opiner: A search and summarization engine for {API} reviews.
\newblock In {\em Proc. 32nd IEEE/ACM International Conference on Automated
  Software Engineering}, page~6, 2017.

\bibitem{Uddin-OpinionValue-TSE2019}
G.~Uddin and F.~Khomh.
\newblock Automatic opinion mining from {API} reviews from stack overflow.
\newblock {\em IEEE Transactions on Software Engineering}, page~35, 2019.

\bibitem{Uddin-OpinerAPIUsageScenario-TOSEM2020}
G.~Uddin, F.~Khomh, and C.~K. Roy.
\newblock Automatic api usage scenario documentation from technical q\&a sites.
\newblock {\em ACM Transactions on Software Engineering and Methodology},
  page~43, 2020.

\bibitem{Uddin-OpinerAPIUsageScenario-IST2020}
G.~Uddin, F.~Khomh, and C.~K. Roy.
\newblock Automatic mining of api usage scenarios from stack overflow.
\newblock {\em Information and Software Technology (IST)}, page~16, 2020.

\bibitem{Uddin-HowAPIDocumentationFails-IEEESW2015}
G.~Uddin and M.~P. Robillard.
\newblock How api documentation fails.
\newblock {\em IEEE Softawre}, 32(4):76--83, 2015.

\bibitem{verdecchia2020architectural}
R.~Verdecchia, P.~Kruchten, and P.~Lago.
\newblock Architectural technical debt: A grounded theory.
\newblock In {\em European Conference on Software Architecture}, pages
  202--219. Springer, 2020.

\bibitem{verdecchia2021building}
R.~Verdecchia, P.~Kruchten, P.~Lago, and I.~Malavolta.
\newblock Building and evaluating a theory of architectural technical debt in
  software-intensive systems.
\newblock {\em Journal of Systems and Software}, 176:110925, 2021.

\bibitem{spectralCluster2}
U.~Von~Luxburg.
\newblock A tutorial on spectral clustering.
\newblock {\em Statistics and computing}, 17(4):395--416, 2007.

\bibitem{wattanakriengkrai2018identifying}
S.~Wattanakriengkrai, R.~Maipradit, H.~Hata, M.~Choetkiertikul, T.~Sunetnanta,
  and K.~Matsumoto.
\newblock Identifying design and requirement self-admitted technical debt using
  n-gram idf.
\newblock In {\em 2018 9th International Workshop on Empirical Software
  Engineering in Practice (IWESEP)}, pages 7--12. IEEE, 2018.

\bibitem{xavier2020beyond}
L.~Xavier, F.~Ferreira, R.~Brito, and M.~T. Valente.
\newblock Beyond the code: Mining self-admitted technical debt in issue tracker
  systems.
\newblock In {\em Proceedings of the 17th International Conference on Mining
  Software Repositories}, pages 137--146, 2020.

\bibitem{synonym_replace_text_aug_2}
R.~Xiang, E.~Chersoni, Y.~Long, Q.~Lu, and C.-R. Huang.
\newblock Lexical data augmentation for text classification in deep learning.
\newblock In {\em Canadian Conference on Artificial Intelligence}, pages
  521--527. Springer, 2020.

\bibitem{xiao2016identifying}
L.~Xiao, Y.~Cai, R.~Kazman, R.~Mo, and Q.~Feng.
\newblock Identifying and quantifying architectural debt.
\newblock In {\em 2016 IEEE/ACM 38th International Conference on Software
  Engineering (ICSE)}, pages 488--498. IEEE, 2016.

\bibitem{yan2018automating}
M.~Yan, X.~Xia, E.~Shihab, D.~Lo, J.~Yin, and X.~Yang.
\newblock Automating change-level self-admitted technical debt determination.
\newblock {\em IEEE Transactions on Software Engineering}, 45(12):1211--1229,
  2018.

\bibitem{yu2020identifying}
Z.~Yu, F.~M. Fahid, H.~Tu, and T.~Menzies.
\newblock Identifying self-admitted technical debts with jitterbug: A two-step
  approach.
\newblock {\em IEEE Transactions on Software Engineering}, 2020.

\bibitem{code_design_relation_2}
N.~Zazworka, M.~A. Shaw, F.~Shull, and C.~Seaman.
\newblock Investigating the impact of design debt on software quality.
\newblock In {\em Proceedings of the 2nd Workshop on Managing Technical Debt},
  pages 17--23, 2011.

\bibitem{zazworka2011investigating}
N.~Zazworka, M.~A. Shaw, F.~Shull, and C.~Seaman.
\newblock Investigating the impact of design debt on software quality.
\newblock In {\em Proceedings of the 2nd Workshop on Managing Technical Debt},
  pages 17--23, 2011.

\bibitem{zazworka2013case}
N.~Zazworka, R.~O. Sp{\'\i}nola, A.~Vetro', F.~Shull, and C.~Seaman.
\newblock A case study on effectively identifying technical debt.
\newblock In {\em Proceedings of the 17th International Conference on
  Evaluation and Assessment in Software Engineering}, pages 42--47, 2013.

\bibitem{zazworka2014comparing}
N.~Zazworka, A.~Vetro, C.~Izurieta, S.~Wong, Y.~Cai, C.~Seaman, and F.~Shull.
\newblock Comparing four approaches for technical debt identification.
\newblock {\em Software Quality Journal}, 22(3):403--426, 2014.

\bibitem{zhang2008text}
W.~Zhang, T.~Yoshida, and X.~Tang.
\newblock Text classification based on multi-word with support vector machine.
\newblock {\em Knowledge-Based Systems}, 21(8):879--886, 2008.

\end{thebibliography}


\begin{thebibliography}{10}

\bibitem{website:javadocse7}
{\em Javadoc SE 7}.
\newblock \url{https://docs.oracle.com/javase/7/docs/api/}, 2020.

\bibitem{Abidi-AntiPatternMultiLanguage-EuroPLoP2019}
M.~Abidi, M.~Grichi, F.~Khomh, and Y.~G. Gu\'{e}h\'{e}neuc.
\newblock Anti-patterns for multi-language systems.
\newblock In {\em 24th European Conference on Pattern Languages of Programs},
  page Article No. 42, 2019.

\bibitem{Abidi-CodeSmellsMultiLanguage-EuroPLoP2019}
M.~Abidi, M.~Grichi, F.~Khomh, and Y.~G. Gu\'{e}h\'{e}neuc.
\newblock Code smells for multi-language systems.
\newblock In {\em 24th European Conference on Pattern Languages of Programs},
  page Article No. 12, 2019.

\bibitem{bert_success_4}
A.~Adhikari, A.~Ram, R.~Tang, and J.~Lin.
\newblock Docbert: Bert for document classification.
\newblock {\em arXiv preprint arXiv:1904.08398}, 2019.

\bibitem{Aghajani-AndroidDocumentation-TSE2019}
E.~Aghajani, G.~Bavota, M.~Linares-V\'{a}squez, and M.~Lanza.
\newblock Automated documentation of android apps.
\newblock {\em IEEE Transactions on Software Engineering}, page~17, 2019.

\bibitem{Aghajani-SoftwareDocPractitioner-ICSE2020}
E.~Aghajani, C.~Nagy, M.~Linares-V\'{a}squez, L.~Moreno, G.~Bavota, M.~Lanza,
  and D.~C. Shepherd.
\newblock Software documentation: The practitioners' perspective.
\newblock In {\em 42nd International Conference on Software Engineering},
  page~12, 2020.

\bibitem{Aghajani-SoftwareDocIssueUnveiled-ICSE2019}
E.~Aghajani, C.~Nagy, O.~L. Vega-M\'{a}rquez, M.~Linares-V\'{a}squez,
  L.~Moreno, G.~Bavota, and M.~Lanza.
\newblock Software documentation issues unveiled.
\newblock In {\em 41st International Conference on Software Engineering}, page
  1199–1210, 2019.

\bibitem{mlknn_MultilabelTextClassificationUsingSemanticFeatures}
W.~Alkhatib, C.~Rensing, and J.~Silberbauer.
\newblock Multi-label text classification using semantic features and
  dimensionality reduction with autoencoders.
\newblock In {\em International Conference on Language, Data and Knowledge},
  pages 380--394. Springer, 2017.

\bibitem{permutation_feature_importance_paper_1}
L.~Breiman.
\newblock Random forests.
\newblock {\em Machine Learning}, 45(1):5--32, 2001.

\bibitem{jargon_paper_1}
O.~M. Bullock, D.~Col{\'o}n~Amill, H.~C. Shulman, and G.~N. Dixon.
\newblock Jargon as a barrier to effective science communication: Evidence from
  metacognition.
\newblock {\em Public Understanding of Science}, 28(7):845--853, 2019.

\bibitem{Cai-FrameworkDocumentation-PhDThesis2000}
I.~Cai.
\newblock {\em Framework Documentation: How to document object-oriented
  frameworks. An Empirical Study}.
\newblock {PhD} in {C}omputer {S}science, University of Illinois at
  Urbana-Champaign, 2000.

\bibitem{Campbell-DeficientDocumentationDetection-MSR2013}
J.~C. Campbell, C.~Zhang, Z.~Xu, A.~Hindle, and J.~Miller.
\newblock Deficient documentation detection: {A} methodology to locate
  deficient project documentation using topic analysis.
\newblock In {\em Proceedings of the 10th International Working Conference on
  Mining Software Repositories}, pages 57--60, 2013.

\bibitem{Carroll-MinimalManual-JournalHCI1987a}
J.~M. Carroll, P.~L. Smith-Kerker, J.~R. Ford, and S.~A. Mazur-Rimetz.
\newblock The minimal manual.
\newblock {\em Journal of Human-Computer Interaction}, 3(2):123--153, 1987.

\bibitem{SVM_SupportVectorNetworks}
C.~Cortes and V.~Vapnik.
\newblock Support-vector networks.
\newblock {\em Machine learning}, 20(3):273--297, 1995.

\bibitem{phi_coefficient_paper}
H.~Cram{\'e}r.
\newblock {\em Mathematical methods of statistics}, volume~43.
\newblock Princeton university press, 1999.

\bibitem{Dagenais-DeveloperLearningResources-PhDThesis2012}
B.~Dagenais.
\newblock {\em Analysis and Recommendations for Developer Learning Resources}.
\newblock {PhD} in {C}omputer {S}science, McGill University, 2012.

\bibitem{Dagenais-TraceabilityLinksRecommendDocumentationEvolution-TSE2014}
B.~Dagenais and M.~P. Robillard.
\newblock Using traceability links to recommend adaptive changes for
  documentation evolution.
\newblock {\em IEEE Transactions on Software Engineering}, 40(11):1126--1146,
  2014.

\bibitem{multilabel_decomp_3_ATutorialOnMultilabelClassification}
A.~C. de~Carvalho and A.~A. Freitas.
\newblock A tutorial on multi-label classification techniques.
\newblock In {\em Foundations of computational intelligence volume 5}, pages
  177--195. Springer, 2009.

\bibitem{DeSouza-DocumentationEssentialForSoftwareMaintenance-SIGDOC2005}
S.~C.~B. de~Souza, N.~Anquetil, and K.~M. de~Oliveira.
\newblock A study of the documentation essential to software maintenance.
\newblock In {\em 23rd annual international conference on Design of
  communication: documenting \& designing for pervasive information}, pages
  68--75, 2005.

\bibitem{Delfim-RedocummentingAPIsCrowdKnowledge-JournalBrazilian2016}
F.~Delfim and M.~M. Kl\'{e}risson Paix\~{a}o, Damien~Cassou.
\newblock Redocumenting apis with crowd knowledge: a coverage analysis based on
  question types.
\newblock {\em Journal of the Brazilian Computer Society}, 29(1), 2016.

\bibitem{bert_base_BertPretrainingOfDeepBidirectionalTransformers}
J.~Devlin, M.-W. Chang, K.~Lee, and K.~Toutanova.
\newblock Bert: Pre-training of deep bidirectional transformers for language
  understanding.
\newblock {\em arXiv preprint arXiv:1810.04805}, 2018.

\bibitem{svm_text_classification_1}
S.~Dumais et~al.
\newblock Using svms for text categorization.
\newblock {\em IEEE Intelligent Systems}, 13(4):21--23, 1998.

\bibitem{SVM_multi_1_AKernelMethodForMultilabelled}
A.~Elisseeff and J.~Weston.
\newblock A kernel method for multi-labelled classification.
\newblock In {\em Advances in neural information processing systems}, pages
  681--687, 2002.

\bibitem{permutation_feature_importance_paper_2}
A.~Fisher, C.~Rudin, and F.~Dominici.
\newblock All models are wrong, but many are useful: Learning a variable's
  importance by studying an entire class of prediction models simultaneously.
\newblock {\em Journal of Machine Learning Research}, 20(177):1--81, 2019.

\bibitem{flesch_readability_paper}
R.~Flesch and A.~J. Gould.
\newblock {\em The art of readable writing}, volume~8.
\newblock Harper New York, 1949.

\bibitem{Forward-RelevanceSoftwareDocumentationTools-DocEng2002}
A.~Forward and T.~C. Lethbridge.
\newblock The relevance of software documentation, tools and technologies: A
  survey.
\newblock In {\em Proc. ACM Symposium on Document Engineering}, pages 26--33,
  2002.

\bibitem{Garousi-UsageUsefulnessSoftwareDoc-IST2015}
G.~Garousi, ahid Garousi-Yusifo\'{g}lu, G.~Ruhe, J.~Zhi, M.~Moussavi, and
  B.~Smith.
\newblock Usage and usefulness of technical software documentation: An
  industrial case study.
\newblock {\em Information and Software Technology}, 57:664--682, 2015.

\bibitem{svm_text_classification_2}
T.~F. Gharib, M.~B. Habib, and Z.~T. Fayed.
\newblock Arabic text classification using support vector machines.
\newblock {\em Int. J. Comput. Their Appl.}, 16(4):192--199, 2009.

\bibitem{bert_success_2}
S.~Gonz{\'a}lez-Carvajal and E.~C. Garrido-Merch{\'a}n.
\newblock Comparing bert against traditional machine learning text
  classification.
\newblock {\em arXiv preprint arXiv:2005.13012}, 2020.

\bibitem{bilstm_FramewisePhonemeClassificationWithBiLSTM}
A.~Graves and J.~Schmidhuber.
\newblock Framewise phoneme classification with bidirectional lstm and other
  neural network architectures.
\newblock {\em Neural networks}, 18(5-6):602--610, 2005.

\bibitem{label_correlation_paper_1}
Q.~Gu, Z.~Li, and J.~Han.
\newblock Correlated multi-label feature selection.
\newblock In {\em Proceedings of the 20th ACM international conference on
  Information and knowledge management}, pages 1087--1096, 2011.

\bibitem{bag_of_words_paper_1}
Z.~S. Harris.
\newblock Distributional structure.
\newblock {\em Word}, 10(2-3):146--162, 1954.

\bibitem{svm_rbf_2_APracticalGuideToSVM}
C.-W. Hsu, C.-C. Chang, C.-J. Lin, et~al.
\newblock A practical guide to support vector classification, 2003.

\bibitem{Jiau-FacingInequalityCrowdSourcedDocumentation-SENOTE2012}
H.~Jiau and F.-P. Yang.
\newblock Facing up to the inequality of crowdsourced api documentation.
\newblock {\em {ACM} SIGSOFT Software Engineering Notes}, 37(1):1--9, 2012.

\bibitem{SVM_multi_2_TextCategorizationWithSVM}
T.~Joachims.
\newblock Text categorization with support vector machines: Learning with many
  relevant features.
\newblock In {\em European conference on machine learning}, pages 137--142.
  Springer, 1998.

\bibitem{Kavaler-APIsUsedinAndroidMarket-SOCINFO2013}
D.~Kavaler, D.~Posnett, C.~Gibler, H.~Chen, P.~Devanbu, and V.~Filkov.
\newblock Using and asking: Apis used in the android market and asked about in
  stackoverflow.
\newblock In {\em In Proceedings of the INTERNATIONAL CONFERENCE ON SOCIAL
  INFORMATICS}, pages 405--418, 2013.

\bibitem{adam_optimizer}
D.~P. Kingma and J.~Ba.
\newblock Adam: A method for stochastic optimization.
\newblock {\em arXiv preprint arXiv:1412.6980}, 2014.

\bibitem{edit_distance_levenshtein_paper}
V.~I. Levenshtein.
\newblock Binary codes capable of correcting deletions, insertions, and
  reversals.
\newblock In {\em Soviet physics doklady}, volume~10, pages 707--710, 1966.

\bibitem{bert_success_5}
X.~Li, L.~Bing, W.~Zhang, and W.~Lam.
\newblock Exploiting bert for end-to-end aspect-based sentiment analysis.
\newblock {\em arXiv preprint arXiv:1910.00883}, 2019.

\bibitem{bert_success_3}
Y.~Liu.
\newblock Fine-tune bert for extractive summarization.
\newblock {\em arXiv preprint arXiv:1903.10318}, 2019.

\bibitem{adam_w}
I.~Loshchilov and F.~Hutter.
\newblock Decoupled weight decay regularization.
\newblock {\em arXiv preprint arXiv:1711.05101}, 2017.

\bibitem{Meij-AssessmentMinimalistApproachDocumentation-SIGDOC1992}
H.~V.~D. Maij.
\newblock A critical assessment of the minimalist approach to documentation.
\newblock In {\em Proc. 10th ACM SIGDOC International Conference on Systems
  Documentation}, pages 7--17, 1992.

\bibitem{Manning-IRIntroBook-Cambridge2009}
C.~D. Manning, P.~Raghavan, and H.~Sch\"{u}tze.
\newblock {\em An Introduction to Information Retrieval}.
\newblock Cambridge Uni Press, 2009.

\bibitem{McBurney-DocumentationSourceCodeSummarization-ICPC2014}
P.~W. McBurney and C.~McMillan.
\newblock Automatic documentation generation via source code summarization of
  method context.
\newblock In {\em 22nd International Conference on Program Comprehension},
  pages 279 -- 290, 2014.

\bibitem{cohen_kappa_paper}
M.~L. McHugh.
\newblock Interrater reliability: the kappa statistic.
\newblock {\em Biochemia medica: Biochemia medica}, 22(3):276--282, 2012.

\bibitem{bag_of_words_for_text_2}
M.~McTear, Z.~Callejas, and D.~Griol.
\newblock Spoken language understanding.
\newblock In {\em The Conversational Interface}, pages 161--185. Springer
  International Publishing, 2016.

\bibitem{Moreno-NLPJavaClasses-ICPC2013}
L.~Moreno, J.~Aponte, G.~Sridhara, A.~Marcus, L.~Pollock, and K.~Vijay-Shanker.
\newblock Automatic generation of natural language summaries for {Java}
  classes.
\newblock In {\em Proceedings of the 21st IEEE International Conference on
  Program Comprehension}, pages 23--32, 2013.

\bibitem{bert_success_6}
M.~Munikar, S.~Shakya, and A.~Shrestha.
\newblock Fine-grained sentiment classification using bert.
\newblock In {\em 2019 Artificial Intelligence for Transforming Business and
  Society (AITB)}, volume~1, pages 1--5. IEEE, 2019.

\bibitem{Nykaza-ProgrammersNeedsAssessmentSDKDoc-SIGDOC2002}
J.~Nykaza, R.~Messinger, F.~Boehme, C.~L. Norman, M.~Mace, and M.~Gordon.
\newblock What programmers really want: Results of a needs assessment for {SDK}
  documentation.
\newblock In {\em Proc. 20th Annual International Conference on Computer
  Documentation}, pages 133--141, 2002.

\bibitem{Parnin-MeasuringAPIDocumentationWeb-Web2SE2011}
C.~Parnin and C.~Treude.
\newblock Measuring api documentation on the web.
\newblock In {\em Proceedings of the 2nd International Workshop on Web 2.0 for
  Software Engineering}, pages 25--30, 2011.

\bibitem{stratified_cross_validation_paper}
V.~L. Parsons.
\newblock Stratified sampling.
\newblock {\em Wiley StatsRef: Statistics Reference Online}, pages 1--11, 2014.

\bibitem{bert_success_7}
Y.~Peng, S.~Yan, and Z.~Lu.
\newblock Transfer learning in biomedical natural language processing: An
  evaluation of bert and elmo on ten benchmarking datasets.
\newblock {\em arXiv preprint arXiv:1906.05474}, 2019.

\bibitem{Glove_GlobalVectorsForWordRepresentation}
J.~Pennington, R.~Socher, and C.~D. Manning.
\newblock Glove: Global vectors for word representation.
\newblock In {\em Proceedings of the 2014 conference on empirical methods in
  natural language processing (EMNLP)}, pages 1532--1543, 2014.

\bibitem{Ponzanelli-PrompterRecommender-EMSE2014}
L.~Ponzanelli, G.~Bavota, M.~{Di Penta}, R.~Oliveto, and M.~Lanza.
\newblock Prompter: Turning the {IDE} into a self-confident programming
  assistant.
\newblock {\em Empirical Software Engineering}, 21(5):2190--2231, 2016.

\bibitem{early_stop_bert_2}
L.~Prechelt.
\newblock Automatic early stopping using cross validation: quantifying the
  criteria.
\newblock {\em Neural Networks}, 11(4):761--767, 1998.

\bibitem{early_stop_bert_1}
L.~Prechelt.
\newblock Early stopping-but when?
\newblock In {\em Neural Networks: Tricks of the trade}, pages 55--69.
  Springer, 1998.

\bibitem{rabbi2020detecting}
F.~Rabbi and M.~S. Siddik.
\newblock Detecting code comment inconsistency using siamese recurrent network.
\newblock In {\em Proceedings of the 28th International Conference on Program
  Comprehension}, pages 371--375, 2020.

\bibitem{CC_2_ClassifierChainsForMultilabel}
J.~Read, B.~Pfahringer, G.~Holmes, and E.~Frank.
\newblock Classifier chains for multi-label classification.
\newblock In {\em Joint European Conference on Machine Learning and Knowledge
  Discovery in Databases}, pages 254--269. Springer, 2009.

\bibitem{traditional_cross_validation_paper}
P.~Refaeilzadeh, L.~Tang, and H.~Liu.
\newblock Cross-validation.
\newblock {\em Encyclopedia of database systems}, 5:532--538, 2009.

\bibitem{Ren-DemystifyOfficialAPIUsageDirectivesWithCorwdExample-ICSE2020}
X.~Ren, J.~Sun, Z.~Xing, X.~Xia, and J.~Sun.
\newblock Demystify official api usage directives with crowdsourced apimisuse
  scenarios, erroneous code examples and patches.
\newblock In {\em 42nd International Conference on Software Engineering},
  page~12, 2020.

\bibitem{Robillard-APIsHardtoLearn-IEEESoftware2009a}
M.~P. Robillard.
\newblock What makes {APIs} hard to learn? {Answers} from developers.
\newblock {\em IEEE Software}, 26(6):26--34, 2009.

\bibitem{Robillard-FieldStudyAPILearningObstacles-SpringerEmpirical2011a}
M.~P. Robillard and R.~DeLine.
\newblock A field study of {API} learning obstacles.
\newblock {\em Empirical Software Engineering}, 16(6):703--732, 2011.

\bibitem{binary_cross_AreLossFunctionSame}
L.~Rosasco, E.~D. Vito, A.~Caponnetto, M.~Piana, and A.~Verri.
\newblock Are loss functions all the same?
\newblock {\em Neural Computation}, 16(5):1063--1076, 2004.

\bibitem{Rossen-SmallTalkMinimalistInstruction-CHI1990a}
M.~B. Rosson, J.~M. Carrol, and R.~K. Bellamy.
\newblock Smalltalk scaffolding: a case study of minimalist instruction.
\newblock In {\em Proc. ACM SIGCHI Conf. on Human Factors in Computing
  Systems}, pages 423--430, 1990.

\bibitem{Haiduc:Summarization}
A.~M. S.~Haiduc, J.~Aponte.
\newblock Supporting program comprehension with source code summarization.
\newblock In {\em In Proceedings of the 32nd International Conference on
  Software Engineering}, pages 223--226, 2010.

\bibitem{bag_of_words_for_text_1}
F.~Sebastiani.
\newblock Machine learning in automated text categorization.
\newblock {\em ACM computing surveys (CSUR)}, 34(1):1--47, 2002.

\bibitem{iterative_stratified_cross_valid}
K.~Sechidis, G.~Tsoumakas, and I.~Vlahavas.
\newblock On the stratification of multi-label data.
\newblock In {\em Joint European Conference on Machine Learning and Knowledge
  Discovery in Databases}, pages 145--158. Springer, 2011.

\bibitem{Shull-InvestigatingReadingTechniquesForOOFramework-TSE2000}
F.~Shull, F.~Lanubile, and V.~R. Basili.
\newblock Investigating reading techniques for object-oriented framework
  learning.
\newblock {\em IEEE Transactions on Software Engineering}, 26(11):1101--1118,
  2000.

\bibitem{Souza-CookbookAPI-BSSE2014}
L.~Souza, E.~Campos, , and M.~Maia.
\newblock On the extraction of cookbooks for apis from the crowd knowledge.
\newblock In {\em Proceedings of the 28th Brazilian Symposium on Software
  Engineering}, pages 21--30, 2014.

\bibitem{Souza-BootstrapAPICodeBookSO-IST2019}
L.~B. Souza, E.~C. Campos, F.~Madeiral, K.~P. {a}o, A.~M. Rocha, and
  M.~de~Almeida~Maia.
\newblock Bootstrapping cookbooks for apis from crowd knowledge on stack
  overflow.
\newblock {\em Information and Software Technology}, 111:3749, 2019.

\bibitem{Sridhara-SummaryCommentsJavaClasses-ASE2010}
G.~Sridhara, E.~Hill, D.~Muppaneni, L.~Pollock, and K.~Vijay-Shanker.
\newblock Towards automatically generating summary comments for java methods.
\newblock In {\em Proceedings of the IEEE/ACM international conference on
  Automated software engineering}, pages 43--52, 2010.

\bibitem{Subramanian-LiveAPIDocumentation-ICSE2014}
S.~Subramanian, L.~Inozemtseva, and R.~Holmes.
\newblock Live {API} documentation.
\newblock In {\em Proceedings of 36th International Conference on Software
  Engineering}, pages 643--652, 2014.

\bibitem{bert_fine_tuning}
C.~Sun, X.~Qiu, Y.~Xu, and X.~Huang.
\newblock How to fine-tune bert for text classification?
\newblock In {\em China National Conference on Chinese Computational
  Linguistics}, pages 194--206. Springer, 2019.

\bibitem{Sunshine-APIProtocolUsability-ICPC2015}
J.~Sunshine, J.~D. Herbsleb, , and J.~Aldrich.
\newblock Searching the state space: A qualitative study of api protocol
  usability.
\newblock In {\em Proceedings of the International Conference on Program
  Comprehension}, pages 82--93, 2015.

\bibitem{Tan-tCommentCodeCommentInconsistency-ICSTVV2012}
S.~H. Tan, D.~Marinov, L.~Tan, and G.~T. Leavens.
\newblock \@tcomment: Testing javadoc comments to detect comment-code
  inconsistencies.
\newblock In {\em International Conference on Software Testing, Verification,
  and Validation}, pages 260 -- 269, 2012.

\bibitem{bert_success_1}
I.~Tenney, D.~Das, and E.~Pavlick.
\newblock Bert rediscovers the classical nlp pipeline.
\newblock {\em arXiv preprint arXiv:1905.05950}, 2019.

\bibitem{Treude-DocumentationQuality-FSE2020}
C.~Treude, J.~Middleton, and T.~Atapattu.
\newblock Beyond accuracy: Assessing software documentation quality.
\newblock In {\em ACM Joint European Software Engineering Conference and
  Symposium on the Foundations of Software Engineering - Vision and Reflections
  Track}, page~4, 2020.

\bibitem{Treude-APIInsight-ICSE2016}
C.~Treude and M.~P. Robillard.
\newblock Augmenting api documentation with insights from stack overflow.
\newblock In {\em Proc. IEEE 38th International Conference on Software
  Engineering}, pages 392--402, 2016.

\bibitem{multilabel_decomp_1_MultilabelClassificationAnOverview}
G.~Tsoumakas and I.~Katakis.
\newblock Multi-label classification: An overview.
\newblock {\em International Journal of Data Warehousing and Mining (IJDWM)},
  3(3):1--13, 2007.

\bibitem{multilabel_decomp_4_LearningFromMultilabelData}
G.~Tsoumakas and M.-L. Zhang.
\newblock Learning from multi-label data.
\newblock 2009.

\bibitem{Uddin-HowAPIDocumentationFails-IEEESW2015}
G.~Uddin and M.~P. Robillard.
\newblock How {API} documentation fails.
\newblock {\em IEEE Softawre}, 32(4):76--83, 2015.

\bibitem{Wang-APIsUsageObstacles-MSR2013}
W.~Wang and M.~W. Godfrey.
\newblock Detecting api usage obstacles: A study of ios and android developer
  questions.
\newblock In {\em In Proceedings of the 10th Working Conference on Mining
  Software Repositories}, pages 61--64, 2013.

\bibitem{Hsu-PracticalSVM-Misc2010}
C.~wei Hsu, C.~chung Chang, and C.~jen Lin.
\newblock A practical guide to support vector classification.

\bibitem{Wen-CodeCommentInconsistencyEmpirical-ICPC2019}
F.~Wen, C.~Nagy, G.~Bavota, and M.~Lanza.
\newblock A large-scale empirical study on code-comment inconsistencies.
\newblock In {\em 27th International Conference on Program Comprehension}, page
  53–64, 2019.

\bibitem{Yang-QueryToUsableCode-MSR2016}
D.~Yang, A.~Hussain, and C.~V. Lopes.
\newblock From query to usable code: an analysis of stack overflow code
  snippets.
\newblock In {\em In Proceedings of the 13th International Conference on Mining
  Software Repositories}, pages 391--402, 2016.

\bibitem{svm_linear_kernel_paper_1}
Y.~Yang and X.~Liu.
\newblock A re-examination of text categorization methods.
\newblock In {\em Proceedings of the 22nd annual international ACM SIGIR
  conference on Research and development in information retrieval}, pages
  42--49, 1999.

\bibitem{svm_rbf_1_AComparisonStudyOfDifferentKernelFunctions}
B.~Yekkehkhany, A.~Safari, S.~Homayouni, and M.~Hasanlou.
\newblock A comparison study of different kernel functions for svm-based
  classification of multi-temporal polarimetry sar data.
\newblock {\em The International Archives of Photogrammetry, Remote Sensing and
  Spatial Information Sciences}, 40(2):281, 2014.

\bibitem{MLkNN_ALazyLearningApproach}
M.-L. Zhang and Z.-H. Zhou.
\newblock Ml-knn: A lazy learning approach to multi-label learning.
\newblock {\em Pattern recognition}, 40(7):2038--2048, 2007.

\bibitem{multilabel_decomp_2_AReviewOnMultilabelLearning}
M.-L. Zhang and Z.-H. Zhou.
\newblock A review on multi-label learning algorithms.
\newblock {\em IEEE transactions on knowledge and data engineering},
  26(8):1819--1837, 2013.

\bibitem{svm_linear_kernel_paper_2}
W.~Zhang, T.~Yoshida, and X.~Tang.
\newblock Text classification based on multi-word with support vector machine.
\newblock {\em Knowledge-Based Systems}, 21(8):879--886, 2008.

\bibitem{Zhia-CostBenefitSoftwareDoc-JSS2015}
J.~Zhia, V.~Garousi-Yusifo\'{g}lubc, B.~Sun, G.~Garousi, S.~Shahnewaz, and
  G.~Ruhe.
\newblock Cost, benefits and quality of software development documentation: A
  systematic mapping.
\newblock {\em Journal of Systems and Software}, 99:175--198, 2015.

\bibitem{Zhou-DocumentationCodeToDetectDirectiveDefects-ICSE2017}
Y.~Zhou, R.~Gu, T.~Chen, Z.~Huang, S.~Panichella, and H.~Gall.
\newblock Analyzing apis documentation and code to detect directive defects.
\newblock In {\em 39th International Conference on Software Engineering}, pages
  27 -- 37, 2017.

\end{thebibliography}
\end{small}

\end{document}